\documentclass[nonblindrev]{informs3} 

\DoubleSpacedXI 

\usepackage{endnotes}
\let\footnote=\endnote

\usepackage{natbib}
 \bibpunct[, ]{(}{)}{,}{a}{}{,}%
 \def\bibfont{\small}%
\TheoremsNumberedThrough     
\ECRepeatTheorems

\EquationsNumberedThrough    

\usepackage[margin=1in]{geometry} 
\usepackage{color}
\usepackage{amsmath,amsfonts,amssymb}
\usepackage{algorithm}
\usepackage{algpseudocode}
\usepackage{bbm}
\usepackage{booktabs}
\usepackage{multirow}
\usepackage{enumerate}
\usepackage{scalerel}
\usepackage{hyperref}
\hypersetup{colorlinks,
  urlcolor=[rgb]{0.5,0.5,0.5},
            citecolor=blue,
            linkcolor=blue}
\usepackage{pifont}
\usepackage[table]{xcolor}
\usepackage{subcaption}

\newcommand{\cmark}{\ding{51}}%
\newcommand{\xmark}{\ding{55}}%
\DeclareMathOperator{\Tr}{Tr}

\renewcommand*\thesection{\arabic{section}}

\begin{document}



\TITLE{Efficient Exponential Tilting for Portfolio Credit Risk}

\ARTICLEAUTHORS{%
  \AUTHOR{Cheng-Der Fuh}
\AFF{Fanhai International School of Finance, Fudan University, Shanghai, China, \EMAIL{cdfu@fudan.edu.cn}} 
\AUTHOR{Chuan-Ju Wang}
\AFF{Research Center for Information Technology Innovation, Academia Sinica, Taipei, Taiwan, \EMAIL{cjwang@citi.sinica.edu.tw}}
}

\ABSTRACT{%
  This paper considers the problem of measuring the credit risk in portfolios of
loans, bonds, and other instruments subject to possible default under
multi-factor models.
Due to the amount of the portfolio, the heterogeneous effect of obligors, and
the phenomena that default events are rare and mutually dependent, it is
difficult to calculate portfolio credit risk either by means of direct analysis
or crude Monte Carlo under such models.
To capture the extreme dependence among obligors, we provide an efficient
simulation method for multi-factor models with a normal mixture copula that
allows the multivariate defaults to have an asymmetric distribution, while most
of the literature focuses on simulating one-dimensional cases.
To this end, we first propose a general account of an importance sampling
algorithm based on an unconventional exponential embedding, which is related to
the classical sufficient statistic.
Note that this innovative tilting device is more suitable for the
multivariate normal mixture model than traditional one-parameter tilting methods
and is of independent interest.
Next, by utilizing a fast computational method for how the rare event occurs
and the proposed importance sampling method, we provide an efficient simulation
algorithm to estimate the probability that the portfolio incurs large losses
under the normal mixture copula. Here the proposed simulation device is based
on importance sampling for a joint probability other than the conditional
probability used in previous studies.
Theoretical investigations and simulation studies, which include an empirical
example, are given to illustrate the method.
}%



\maketitle

%


\section{Introduction}

Loss ensuing from the nonfulfillment of an obligor to make required payments,
typically referred to as credit risk, is one of the prime concerns of
financial institutions. 
Modern credit risk management usually takes a portfolio approach to measure and
manage this risk, in which the dependence among sources of credit risk
(obligors) in the portfolio is modeled. 
With the potential default of obligors, the portfolio approach evaluates the
impact of the dependence among them on the probability of multiple defaults and
large losses.
Moreover, the default event triggered by an obligor is generally captured via 
so-called threshold models, in which an obligor defaults when a latent variable
associated with this obligor goes beyond (or falls below) a predetermined
threshold. 
This important concept is shared among all models originating from Merton's
credit risk model~\citep{merton1974pricing}; the latent variable associated
with each obligor is modeled using multiple factors occurring as a result of
factor analysis, and thus captures common macroeconomic or industry-specific
effects.

To model the dependence structure that shapes the multivariate default
distribution, {\it copula} functions have been widely adopted in
the literature~\citep[e.g.,][]{Li2000,glasserman2005importance,glasserman2007large,glasserman2008fast,bassamboo2008portfolio,
chan2010efficient}.
One of the most widely used models is the normal copula model, which
assumes that the latent variables follow a multivariate normal distribution and
forms the basis of many risk-management systems such as J.\ P.\ Morgan's
CreditMetrics~\citep{gupton1997creditmetrics}.
In spite of its popularity, some empirical studies suggest that strong
dependence exhibited among financial variables is difficult to capture in
the normal copula model~\citep{mashal2002beyond}.
Therefore, in view of this limitation, \cite{bassamboo2008portfolio} present a
$t$-copula model derived from the multivariate $t$-distribution to capture the
relatively strong dependence structure of financial variables.
Further studies can be found in~\cite{chan2010efficient} and~\cite{scott2015general}.


Monte Carlo simulation is the most widely adopted computational technique when
modeling the dependence among sources of credit risk; it however converges 
slowly, especially with low-probability
events. To obtain rare-event probabilities more efficiently in simulation, one common
approach is to shift the factor mean via importance sampling, a type of
variance reduction~\citep[e.g.,][]{AsmussenGlynn2007,rubinstein2011simulation}.
Various importance sampling techniques for rare-event simulation for credit
risk measurement have been studied in the
literature~\citep[e.g.,][]{glasserman2005importance,glasserman2007large,bassamboo2008portfolio,botev2013markov,scott2015general,liu2015simulating}.
For example, \cite{glasserman2005importance} develop a two-step importance
sampling approach for the normal copula model and derive the logarithmic limits
for the tail of the loss distribution associated with single-factor homogeneous
portfolios.
Due to the difficulty of generalizing the approach to the general multi-factor
model, \cite{glasserman2007large} derive the logarithmic asymptotics for the
loss distribution under the multi-factor normal copula model, the results of
which are later utilized in~\cite{glasserman2008fast} to develop importance
sampling techniques to estimate the tail probabilities of large losses, considering
different types of obligors.
Others focus on importance sampling techniques for the
$t$-copula model; for instance, \cite{bassamboo2008portfolio} present two
importance sampling algorithms to estimate the probability of large losses
under the assumption of the multivariate $t$-distribution. 
\cite{chan2010efficient} propose two simple simulation algorithms based on
conditional Monte Carlo, which utilizes the asymptotic description of how the
rare event occurs; later in~\cite{scott2015general}, the authors develop a
novel importance sampling algorithm that requires less computational time with
only slightly less accurate results than that in~\cite{chan2010efficient}.
However, most proposed importance sampling techniques for both
normal copula and $t$-copula models choose tilting parameters by minimizing the
upper bound of the second moment of the importance sampling estimator based on
large deviations theory, and in these papers they only focus on the {\it
one-dimensional} case.\footnote{Although some papers have mentioned that
they can handle $d$-factor models, the factors follow i.i.d. Gaussian
variables and thus can be reduced to the one-dimensional case via Cholesky
decomposition.}

An alternative for choosing a tilting parameter is based on the criterion of
minimizing the variance of the importance sampling estimator directly. For
example, \cite{dh1991} and \cite{FuhHu2004} study efficient simulations for
bootstrapping the sampling distribution of statistics of interest. \cite{SuFu2000}
and \cite{SuFu2002} minimize the variance under the original probability
measure.  \cite{Fuhet2011} apply the importance sampling method for
value-at-risk (VaR) computation under a multivariate $t$-distribution.

Along the line of minimizing the variance of the importance sampling estimator,
in this paper we provide an efficient simulation method for the problem in
portfolio risk under the normal mixture copula model, which includes the
popular normal copula and $t$-copula models as special cases.
Here, we consider grouped normal mixture copulas in our general model setting.
An important idea of the normal mixture model is the inclusion of randomness
into the covariance matrix of a multivariate normal distribution via a set of
positive mixing variables, which could be interpreted as ``shock
variables''~\citep[see][]{bassamboo2008portfolio,mcneil2015quantitative} in the
context of modeling portfolio credit risk.
There have been various empirical and theoretical studies related to
normal mixture models in the
literature~\cite[e.g.,][]{barndorff1978hyperbolic,barndorff1997normal,eberlein1995hyperbolic};
these are popular in financial applications because such
models appear to yield a good fit to financial return data and are consistent
with continuous-time
models~\citep{eberlein1995hyperbolic}.

There are two aspects in this study.
To begin with, based on the criterion of minimizing the variance of the
importance sampling estimator, we propose an innovative importance sampling
algorithm, which is based on an unconventional \emph{sufficient (statistic)
exponential embedding}.
Here we call it sufficient exponential tilting because the form of the
embedding is selected based on the sufficient statistic of the underlying
distribution, for which more than one parameter (usually two parameters) can be tilted in our method;
for instance, the tilting parameters can be the location and scale parameters
in the multivariate normal distribution.
Theoretical investigations and numerical studies are given to support our newly
proposed importance sampling method.
Note that the innovative tilting formula is more suitable for the grouped
normal mixture copula model, and is of independent interest.
It is worth mentioning that for normal, multivariate normal, and Gamma
distributions, our simulation study shows that the sufficient exponential
tilting performs~2 to~5 times better than the classical one-parameter
exponential tilting for some simple rare event simulations.
In particular, when applying the sufficient exponential tilting in the
normal mixture models, the tilting parameter can be either the shape or the
rate parameter for the underlying Gamma distribution, which results in a more
efficient simulation.

Next, by utilizing a fast computational method for how the rare event occurs
and the proposed importance sampling method, we provide an efficient simulation
algorithm for a multi-factor model with the normal mixture copula model
to estimate the probability that the portfolio incurs large losses.
To be more precise, in this stage, we use the inverse Fourier transform to handle
the distribution of total losses, i.e., the sum of $n$ ``weighted''
independent, but non-identically distributed Bernoulli random variables.
An automatic variant of Newton's method is introduced to determine the optimal
tilting parameters. 
Note that the proposed simulation device is based on importance sampling for a
joint probability other than the conditional probability used in previous
studies.
Finally, to illustrate the applicability of our method, we provide numerical
results of the proposed algorithm under various copula settings, and highlight
some insights of the trade-off between the reduced variances and increased
computational time.
Moreover, we also give an empirical example which contains a set of parameters
of a multi-factor model calibrated from data of the CDXIG index.
In particular, our contributions are summarized below:
\begin{enumerate}
  \item
    We propose a general account of an importance sampling algorithm based on
    an innovative sufficient exponential tilting device.  Our proposed
    sufficient exponential embedding essentially differs from the one-parameter
    tilting methods commonly adopted in the literature for both normal copula
    or $t$-copula models and is more suitable for the grouped normal mixture
    copula model.
    Theoretical investigations and numerical studies are given to support our
    method.
  \item
    Based on the proposed tilting method, an efficient simulation method is
    presented for multi-factor models with a normal mixture copula.
    Note that the proposed algorithm is a multi-dimensional method, whereas
    most of the previous literature focuses on simulating one-dimensional
    cases.
  \item
    Extensive simulation studies attest the capability and performance of the
    proposed method.
    The relation between variance reduction factors and time consumption ratios
    suggests that the proposed algorithm achieves good performance
    and thus makes a practical contribution to measuring portfolio credit risk
    in normal mixture copula models.

\end{enumerate}

The remainder of this paper is organized as follows.
Section~\ref{sec:problem} formulates the problem of estimating large portfolio
losses and presents the normal mixture copula model.
Section~\ref{sec:is} presents a general account of importance sampling based on
the sufficient exponential embedding.
We then study the proposed optimal importance sampling for portfolio loss under
the normal mixture copula model in Section~\ref{sec:sampling}.
The performance of our methods is demonstrated with an extensive
simulation study and an empirical example in Section~\ref{sec:results} and
Section~\ref{sec:empirical}, respectively. Section~\ref{sec:conclude} concludes.
The proofs are deferred to the Appendix. 

\section{Portfolio loss under the normal mixture copula model}\label{sec:problem}

Consider a portfolio of loans consisting of $n$ obligors, each of whom has a small probability of default.
We further assume that the loss resulting from the default of the $k$-th obligor,
denoted as $c_k$ (monetary units), is known.
In copula-based credit models, dependence among default indicator for each
obligor is introduced through a vector of latent variables ${
X}=(X_1,\cdots, X_n)$, where the $k$-th obligor defaults if $X_k$ exceeds some chosen threshold $\chi_k$.
The total loss from defaults is then denoted by
\begin{equation}
  \label{eq:loss}
  L_n=c_1\mathbbm{1}_{\{X_1>\chi_1\}}+\cdots+c_n\mathbbm{1}_{\{X_n>\chi_n\}},
\end{equation}
where $\mathbbm{1}$ is the indicator function.
Particularly, the problem of interest is to estimate the probability of losses,
$P(L_n > \tau)$, especially at large values of $\tau$.

As mentioned earlier in the introduction, the widely-used normal copula model
might assign an inadequate probability to the event of many simultaneous defaults
in a portfolio.
In view of this,~\cite{bassamboo2008portfolio,chan2010efficient} set forth the
$t$-copula model for modeling portfolio credit risk.
In this paper, we further consider the normal mixture
model~\citep{mcneil2015quantitative}, including normal copula and $t$-copula
model as special cases, for the generalized $d$-factor model of the form 
\begin{equation}
  \label{eq:Xk}
  X_k=\rho_{k1}\sqrt{W_1}{Z_1}+\cdots+\rho_{kd}\sqrt{W_d}{Z_d}+\rho_k\sqrt{W_{d+1}}{\epsilon_k},\,\,\,\,\, k=1,\ldots, n, 
\end{equation}
in which 
\begin{itemize}
  \item
    ${Z}=(Z_1,\ldots,Z_d)^\intercal$ follows a $d$-dimensional multivariate normal
    distribution with zero mean and covariance matrix $\Sigma$, where$~^\intercal$ denotes vector transpose;
  \item
    ${W}=(W_1,\ldots,W_d,W_{d+1})$ are non-negative scalar-valued random
    variables which are independent of $Z$, and each $W_j$ is a shock variable
    and independent from each other, for $j=1,\ldots,d+1$, or  $W_1 = W_2 = \cdots = W_{d+1}$ is a common gamma random variable;
  \item
   $\epsilon_k\sim N(0,\sigma_\epsilon^2)$ is an idiosyncratic risk associated with the
    $k$-th obligor, for $k=1,\cdots,n$;
  \item
    $\rho_{k1},\ldots,\rho_{kd}$ are the factor loadings for the $k$-th obligor, and
    $\rho_{k1}^2+\cdots+\rho_{kd}^2\leq 1$; 
  \item
    $\rho_k=\sqrt{1-\left(\rho_{k1}^2+\cdots+\rho_{kd}^2\right)}$, for $k=1,\cdots,n$.
\end{itemize}

Model (\ref{eq:Xk}) is the so-called grouped normal mixture copula in \cite{mcneil2015quantitative}.
The above distributions are known as variance mixture models, which are
constructed by drawing randomly from this set of component multivariate normals
based on a set of ``weights'' controlled by the distribution of $W$.
Such distributions enable us to blend in multiplicative shocks via the
variables $W$, which could be interpreted as shocks that arise from new
information.
Note that in the case that $W_1 = W_2 = \cdots = W_{d+1}$ is a common gamma
random variable, $X$ forms a multivairate $t$-distribtion, which is the most
popular form in financial modeling.
In addition, we consider the case that $W_j$, $j=1,\ldots, d+1$, in
Equation~(\ref{eq:Xk}) to have a generalized inverse Gaussian (GIG) distribution. 
The GIG mixing distribution, a special case of the symmetric generalized
hyperbolic (GH) distribution, is very flexible for modeling financial returns.
Moreover, GH distributions also include the symmetric normal inverse Gaussian
(NIG) distribution and a symmetric multivariate distribution with hyperbolic
distribution for its one-dimensional marginal as interesting examples.
Note that this class of distributions has become popular in the financial
modeling literature. An important reason is their link to L\'evy
processes (like Brownian motion or the compound Poisson distribution) that are
used to model price processes in continuous time. 
For example, the generalized hyperbolic distribution has been used to model
financial returns in~\cite{eberlein1995hyperbolic,eberlein1998new}. 
The reader is referred to \citep{mcneil2015quantitative} for more details.

We now define the tail probability of total portfolio losses conditional on
$Z$ and $W$. Specifically, the tail probability of total portfolio
losses conditional on the factors $Z$ and $W$, denoted as $\varrho(Z,W)$ is
defined as
\begin{equation}\label{hzw}
  \varrho({Z},{W})=P(L_n>\tau|({Z},{W})).
\end{equation}
The desired probability of losses can be represented as \begin{equation}\label{ProbLoss}
  P(L_n>\tau) = E\left[\varrho(Z,W)\right].
\end{equation}

For an efficient Monte Carlo simulation of the probability of total
portfolio losses (\ref{ProbLoss}), we apply importance sampling to the
distributions of the factors ${Z} =(Z_1,\ldots,Z_d)^\intercal$ and
${W}=(W_1,\ldots,W_{d+1})^\intercal$ (see Equation~(\ref{eq:Xk})). 
In other words, we attempt to choose importance sampling distributions for both
${Z}$ and ${W}$ that reduce the variance in estimating the integral
$E[\varrho(Z,W)]$ against the original densities of ${Z}$ and ${W}$. 

As noted in~\cite{glasserman2005importance} for normal copula models, the
simulation of (\ref{ProbLoss}) involves two rare events: the default event and
the total portfolio loss event.
For the normal mixture model~(\ref{eq:Xk}), this makes the simulation of
(\ref{ProbLoss}) even more challenging.  For a general simulation algorithm for
this type of problems, we simulate $P(L_n > \tau)$ as expected value of
$\varrho(Z,W)$ in (\ref{ProbLoss}).
Our device is based on a joint probability simulation rather than the
conditional probability simulation considered in the literature.
Moreover, we note that the simulated distributions~-- the multivariate normal
distribution $Z$\footnote{Here we treat the mean vector and variance-covariance
matrix as two parameters.} and the commonly adopted Gamma
distribution for $W$~-- are both two-parameter distributions.
This motivates us to study a sufficient exponential tilting in the next
section.

\section{Sufficient exponential tilting}\label{sec:is}

Let $(\Omega, {\cal F}, P)$ be a given probability space. Let
$X=(X_1,\ldots,X_d)^\intercal$ be a $d$-dimensional random vector having $f(x)
= f(x_1,\ldots,x_d)$ as a probability density function (pdf),
with respect to the Lebesgue measure $\mathcal{L}$, under the probability
measure $P$. 
Let $\wp(\cdot)$ be a real-valued function from $\mathbbm{R}^d$ to $\mathbbm{R}$.
The problem of interest is to calculate the expectation of $\wp(X)$,
\begin{equation}
\label{eqn:mu}
m = E_P[\wp(X)],
\end{equation}
where $E_P[\cdot]$ is the expectation operator under the probability measure $P$. 

To calculate the value of (\ref{eqn:mu}) using importance sampling, one 
selects a sampling probability measure $Q$ under which $X$ has a pdf $q(x) =
q(x_1,\ldots,x_d)$ with respect to the Lebesgue measure $\mathcal{L}$.
The probability measure $Q$ is assumed to be absolutely
continuous with respect to the original probability measure $P$. Therefore,
Equation~(\ref{eqn:mu}) can be written as
\begin{equation}
\label{eqn:IS}
\int_{\mathbbm{R}^d} \wp(x)f(x)dx=\int_{\mathbbm{R}^d} \wp(x) \frac{f(x)}{q(x)}q(x)dx = \mbox{E}_Q\left[\wp(X)\frac{f(X)}{q(X)}\right],
\end{equation}
where $\mbox{E}_Q[\cdot]$ is the expectation operator under which $X$ has a pdf
$q(x)$ with respect to the Lebesgue measure $\mathcal{L}$.
The ratio $f(x)/q(x)$ is called the importance sampling weight, the
likelihood ratio, or the Radon-Nikodym derivative.

Here, we focus on the exponentially tilted probability measure of $P$.
Instead of considering the commonly adopted one-parameter exponential tilting
appeared in the literature~\citep{AsmussenGlynn2007}, we propose a sufficient
exponential tilting algorithm.  
To the best of our knowledge, the use of the sufficient exponential
embedding seems to be new in the literature.
As will be seen in the following examples, the tilting probabilities for
existent two-parameter distributions, such as Gamma distribution and normal
distribution, can be obtained by solving simple formulas.

Let $Q_{\theta,\eta}$ be the tilting probability measure, where the subscript
$\theta=(\theta_1,\ldots,\theta_p)^\intercal \in \Theta \subset \mathbbm{R}^p$
and $\eta=(\eta_1,\cdots,\eta_q)^\intercal \in H \subset \mathbbm{R}^q$ are the
tilting parameters. 
Here $p$ and $q$ denote the number of parameters in $\Theta$ and $H$,
respectively. Let $h_1(x)$ be a function from $\mathbbm{R}^d$ to $\mathbbm{R}^p$, and $h_2(x)$ be a function from $\mathbbm{R}^d$ to $\mathbbm{R}^q$.
Assume that the moment-generating function of $(h_1(X), h_2(X))$ exists and is denoted
by $\Psi(\theta,\eta)$. Let $f_{\theta,\eta}(x)$ be the pdf of $X$ under the
exponentially tilted probability measure $Q_{\theta,\eta}$, defined by
\begin{equation}\label{tilt}
f_{\theta,\eta}(x)=\frac{\mbox{e}^{\theta^\intercal h_1(x) + \eta^\intercal h_2(x)}}{\Psi(\theta,\eta)} f(x)=\mbox{e}^{\theta^\intercal h_1(x) + \eta^\intercal h_2 (x) -\psi(\theta,\eta)} f(x),
\end{equation}
where $\psi(\theta,\eta)=\ln \Psi(\theta,\eta)$ is the cumulant function.
Note that in (\ref{tilt}), we present one type of
parameterization which is suitable for our derivation. Explicit representations
of $h_1(x)$ and $h_2(x)$ for specific distributions are given in the following
examples and remark, which include the normal and multivariate normal
distributions, and the Gamma distribution.

Consider the sufficient exponential embedding. Equation~(\ref{eqn:IS}) becomes
\begin{equation*}
\label{eqn:IS_Qtheta}
\int_{\mathbbm{R}^d} \wp(x)f(x)dx=\int_{\mathbbm{R}^d} \wp(x) \frac{f(x)}{f_{\theta,\eta}(x)}f_{\theta,\eta}(x)dx = E_{Q_{\theta,\eta}}\left[\wp(X)\mbox{e}^{-(\theta^\intercal h_1(X) + \eta^\intercal h_2 (X))+ \psi(\theta,\eta)}\right].
\end{equation*}
Because of the unbiasedness of the importance sampling estimator, its variance is
\begin{equation}
\label{eqn:Var_Qtheta}
Var_{Q_{\theta,\eta}}\left[\wp(X)\mbox{e}^{-(\theta^\intercal h_1(X) + \eta^\intercal h_2 (X))
+ \psi(\theta,\eta)}\right]
= E_{Q_{\theta,\eta}}\left[\bigg(\wp(X)\mbox{e}^{-(\theta^\intercal h_1(X) + \eta^\intercal h_2 (X))+ \psi(\theta,\eta)}\bigg)^2\right] - m^2,
\end{equation}
where $m=\int_{\mathbbm{R}^d} \wp(x)f(x)dx.$
For simplicity, we assume that the variance of the importance sampling estimator exists.

Define the first term of the right-hand side (RHS) of (\ref{eqn:Var_Qtheta}) by
$G(\theta,\eta)$.
Then minimizing $Var_{Q_{\theta,\eta}}\left[\wp(X)\mbox{e}^{-(\theta^\intercal
h_1(X) + \eta^\intercal h_2 (X))+ \psi(\theta,\eta)}\right]$ is equivalent to
minimizing $G(\theta,\eta)$.
Standard algebra gives a simpler form of $G(\theta,\eta)$:
\begin{equation}
\label{eq:G}
G(\theta,\eta):=E_{Q_{\theta,\eta}}\left[\bigg(\wp(X)\mbox{e}^{-(\theta^\intercal h_1(X) + \eta^\intercal h_2 (X))+ \psi(\theta,\eta)}\bigg)^2\right]= E_P\left[\wp^2(X)\mbox{e}^{-(\theta^\intercal h_1(X) + \eta^\intercal h_2(X))+ \psi(\theta,\eta)} \right],
\end{equation}
which is used to find the optimal tilting parameters.
In the following theorem, we show that $G(\theta,\eta)$ in (\ref{eq:G}) is a
convex function in $\theta$ and $\eta$.
This property ensures that there exists no multi-mode problem  in the search
stage when determining the optimal tilting parameters.

To minimize $G(\theta,\eta)$, the first-order condition requires the solution
of $\theta, \eta$, denoted by $\theta^\ast,\eta^\ast$, to satisfy
$\nabla_\theta G(\theta,\eta)\mid_{\theta=\theta^\ast}=0$, and $\nabla_\eta
G(\theta,\eta)\mid_{\eta=\eta^\ast}=0$, where $\nabla_\theta$ denotes the
gradient with respect to $\theta$ and $\nabla_\eta$ denotes the gradient with
respect to $\eta$.
Under the condition that $X$ is in an exponential family, and $\psi(\theta, \eta)$
are bounded continuous differentiable function with respect to $\theta$ and
$\eta$, by the dominated convergence theorem, simple calculation yields
\begin{eqnarray*}
\nabla_\theta G(\theta,\eta) &=& E_{P}\left[\wp^2(X) 
(- h_1(X) + \nabla_\theta \psi(\theta,\eta))\mbox{e}^{-(\theta^\intercal h_1(X) + \eta^\intercal h_2 (X))+ \psi(\theta,\eta)}\right], \\
\nabla_\eta G(\theta,\eta) &=& E_{P}\left[\wp^2(X) 
(-  h_2(X) + \nabla_\eta \psi(\theta,\eta))\mbox{e}^{-(\theta^\intercal h_1(X) + \eta^\intercal h_2(X))+ \psi(\theta,\eta)}\right],
\end{eqnarray*}
and, therefore, $(\theta^\ast, \eta^\ast)$ is the root of the following system
of nonlinear equations,
\begin{eqnarray}
\label{eqn:foc1}
{\nabla_\theta \psi(\theta,\eta)} &=& \frac{E_{P}\left[\wp^2(X)h_1(X) \mbox{e}^{-(\theta^\intercal h_1(X) + \eta^\intercal h_2(X))}\right]}{E_{P}\left[\wp^2(X)\mbox{e}^{-(\theta^\intercal h_1(X) + \eta^\intercal h_2 (X))}\right]}, \\
\label{eqn:foc2}
{\nabla_\eta \psi(\theta,\eta)}& =& \frac{E_{P}\left[\wp^2(X) h_2(X)\mbox{e}^{-(\theta^\intercal h_1(X) + \eta^\intercal h_2(X))}\right]}{E_{P}\left[\wp^2(X)\mbox{e}^{-(\theta^\intercal h_1(X) + \eta^\intercal 
h_2(X))}\right]}.
\end{eqnarray}

To simplify the RHS of (\ref{eqn:foc1}) and (\ref{eqn:foc2}), we
define the conjugate measure $\bar{Q}_{\theta,\eta}:=\bar{Q}^{\wp}_{\theta,\eta}$
of the measure $Q$ with respect to the payoff function $\wp$ as
\begin{equation}
\label{eq:tildeQ}
\frac{d\bar{Q}_{\theta,\eta}}{dP} =\frac{\wp^2(X)\mbox{e}^{-(\theta^\intercal h_1(X) + \eta^\intercal 
h_2(X))}}{E_P[\wp^2(X)\mbox{e}^{-(\theta^\intercal h_1(X) + \eta^\intercal h_2(X))}]} = \wp^2(X)\mbox{e}^{-(\theta^\intercal h_1(X) + \eta^\intercal h_2(X)) - \bar{\psi}(\theta,\eta)},
\end{equation}
where $\bar{\psi}(\theta,\eta)$ is $\log\bar{\Psi}(\theta,\eta)$ with
$\bar{\Psi} (\theta,\eta)=E_P[\wp^2(X)\mbox{e}^{-(\theta^\intercal h_1(X) +
\eta^\intercal h_2(X))}]$. 
Then the RHS of (\ref{eqn:foc1}) equals $E_{\bar{Q}_{\theta,\eta}}[h_1(X)]$, the
expected value of $h_1(X)$ under $\bar{Q}_{\theta,\eta}$, and the RHS of
(\ref{eqn:foc2}) equals $E_{\bar{Q}_{\theta,\eta}}[ h_2(X)]$, the expected value
of $h_2(X)$ under $\bar{Q}_{\theta,\eta}$.

The following theorem establishes the existence, uniqueness, and characterization for
the minimizer of (\ref{eq:G}). Before that, to ensure the finiteness of the
moment-generating function $\Psi(\theta,\eta)$, we add a condition that
$\Psi(\theta,\eta)$ is steep, {cf. \cite{AsmussenGlynn2007}}.
To define steepness, let $\theta^{-}_i= (\theta_1,\cdots, \theta_i,\cdots,\theta_p) \in \Theta$ such
that all $\theta_k$ fixed for $k=1,\cdots,i-1,i+1,\cdots,p$ except the $i$-th component. Denote
$\eta_j^- \in H$ in the same way for $j=1,\cdots,q$. Now, let  $\theta_{i,\max}
:= \sup \{ \theta_i: \Psi(\theta^{-}_i,\eta) < \infty \}$ for $i=1,\cdots,p$,
and $\eta_{j,\max} := \sup \{ \eta_j: \Psi(\theta,\eta^-_j) < \infty \}$ for
$j=1,\cdots,q$ (for light-tailed distributions, we have $0 < \theta_{i,\max}
\leq \infty$ for $i=1,\cdots,p$, and $0 < \eta_{j,\max} \leq \infty$ for
$j=1,\cdots,q$.). Then steepness means $\Psi(\theta,\eta) \to \infty$ as
$\theta_i \to \theta_{i,\max}$ for $i=1,\cdots,p$, or $\eta_j \to
\eta_{j,\max}$ for $j=1,\cdots,q$.

The following conditions are used in Theorem~\ref{thm:theta*}.

\noindent
i) $\bar{\Psi}(\theta,\eta)\Psi(\theta,\eta) \to \infty$  as $\theta_i \to
\theta_{i,\max}$ for $i=1,\cdots,p$, or $\eta_j \to \eta_{j,\max}$ for
$j=1,\cdots,q$;

\noindent
ii) $G(\theta,\eta)$ is a continuous differentiable function on $\Theta \times H$, and
\begin{eqnarray}\label{minimizer}
\sup_{i=1,\cdots,p,~ j = 1,\cdots, q} \bigg\{ \lim_{ \theta_i \to \theta_{i,\max} } \frac{\partial G(\theta,\eta)}{
\partial \theta_i}, \lim_{ \eta_j \to \eta_{j,\max} } \frac{\partial G(\theta,\eta)}{
\partial \eta_j} \bigg\} > 0. 
\end{eqnarray}
{Note that condition i) or ii) is used to guarantee the existence of the minimum point. More details can be found in the Appendix.}

\begin{theorem}
\label{thm:theta*}
Suppose the moment-generating function $\Psi(\theta,\eta)$ of $(h_1(X),h_2(X))$ exists
for $\theta \in \Theta \subset {\mathbbm{R}^p}$ and $\eta \in H \subset
{\mathbbm{R}^q}$.
Assume $\Psi(\theta,\eta)$ is steep. 
Furthermore, assume either i) or ii) holds. 
Then $G(\theta,\eta)$ defined in {\rm (\ref{eq:G})} is a convex function in $(\theta,\eta)$, and there exists a unique minimizer of {\rm (\ref{eq:G})}, which
satisfies
\begin{eqnarray}
\label{eq:theta_ast1}
{\nabla_\theta \psi(\theta,\eta) } &=& E_{\bar{Q}_{\theta,\eta}}[h_1(X)], \\
\label{eq:theta_ast2}
{\nabla_\eta \psi(\theta,\eta) } &=& E_{\bar{Q}_{\theta,\eta}}[ h_2(X)].
\end{eqnarray}
\end{theorem}

The proof of Theorem~\ref{thm:theta*} is given in the Appendix.


\begin{remark}
a) To keep the exponentially tilted probability measure
  $Q_{\theta,\eta}$ within the same exponential family as the original
  probability measure $P$, one possible selection of $h_1(x)$ and $h_2(x)$ is
  based on the sufficient statistic of the original probability distribution. For example, for
  the normal distribution, the sufficient statistic is $T(x)=[x\,\,x^2]$ and
  thus $h_1(x)=x$ and $h_2(x)=x^2$; for the gamma distribution, the sufficient
  statistic is $T(x)=[\log x\,\, x]$ and thus $h_1(x)=\log x$ and $h_2(x)=x$,
  and for the beta distribution, the sufficient statistic is $T(x)=[\log x\,\,
  \log(1-x)]$ and thus $h_1(x)=\log x$ and $h_2(x)=\log(1-x)$.
  Such a device can be applied to other distributions as well, such as the lognormal
  distribution, the inverse Gaussian distribution and the inverse gamma distribution, etc..
  
 b) Now, we provide a heuristic explanation for this device. The idea of using
  sufficient statistic for exponential tilting is
  that we can treat this tilting as a {\it sufficient exponential
  tilting} within the same given parametric family. Furthermore, by
  using the Fisher-Neyman factorization theorem in exponential family, we note that this device
  can provide the maximum degree of freedom for exponential tilting within the
  same exponential family. We also expect to have an analogy parallel to the
  Rao-Blackwell theorem for the sufficient exponential tilting: minimize the mean
  square loss among all possible importance sampling in the same exponential
  embedding. A theoretical study for these properties will be investigated in a
  separate paper.
\end{remark}

To illustrate the proposed sufficient exponential tilting, we here present two
examples: multivariate normal and Gamma distributions.
We choose these two distributions to indicate the location and scale properties
of the sufficient exponential tilting used in our general framework. In these
examples, by using a suitable re-parameterization, we obtain neat tilting
formulas for each distribution based on its sufficient statistic. Our
simulation studies also show that the proposed sufficient exponential tilting
performs 2 to 5 times better than the classical one-parameter exponential
tilting for some simple rare events. 

We here check the validity of applying Theorem~\ref{thm:theta*} for each
example.
First, we note that both multivariate normal distribution and Gamma distribution 
are steep.
Next, it is easy to see that the sufficient conditions
$\bar{\Psi}(\theta)\Psi(\theta) \to \infty$ as $\theta_i \to \theta_{i,\max}$
for $i=1,\cdots,p$, or $\eta_j \to \eta_{j,\max}$ for $j=1,\cdots,q$ in Theorem
\ref{thm:theta*} hold in each example.
For example, when $X\sim N_d(\textbf{0},\Sigma)$, then $\Psi(\theta) = O(e^{\|\theta\|^2})$
approaches $\infty$ sufficiently quickly. Another simple example illustrated here is when $\wp(X) = \textbf{1}_{\{X \in A\}}$ and $A:= [a_1, \infty) \times \cdots \times [a_d,\infty)$, with $a_i > 0$ for all $i=1,\cdots,d$, and $X$ has a $d$-dimensional standard normal
distribution, then it is easy to verify that the sufficient conditions in Theorem~\ref{thm:theta*} hold. 


\begin{example}
  {Multivariate normal distribution.}

To illustrate the concept of the sufficient exponential embedding, we first
consider a one-dimensional normal distributed random variable as an example.
Let $X$ be a random variable with the standard normal distribution, denoted by
$N(0,1)$, with probability density function (pdf)
$\frac{dP}{d\mathcal{L}}=e^{-{x^2}/{2}}/\sqrt{2\pi}.$
By using the sufficient exponential embedding in (\ref{tilt}) with $h_1(x):=x$
and $h_2(x):=x^2$, we have 
\begin{eqnarray}\label{tilt-normal}
  \frac{dQ_{\theta,\eta}/{d\mathcal{L}}}{dP/{d\mathcal{L}}} &=& \frac{\exp\{\theta h_1(x) + \eta h_2(x)\}}{E[\exp\{\theta h_1(X) + \eta h_2(X)\}]} \nonumber \\
&=& \sqrt{1-2 \eta} \exp\{\eta x^2 + \theta x - \theta^2/(2-4\eta)\}.
\end{eqnarray}
In this case, the tilting probability measure $Q_{\theta,\eta}$ is $N(\theta/(1-2\eta),
1/(1-2 \eta))$, with $\eta < 1/2,$ a location-scale family. 

For the event of $\wp(X)=\mathbbm{1}_{\{X > a\}}$ for $a > 0$, define
$\bar{Q}_{\theta,\eta}$ as $N(-\theta/(1+2\eta), 1/(1+2 \eta))$ with $\eta > -1/2$.
For easy presentation, we consider the standard parameterization by letting
$\mu:=\theta/(1-2\eta)$, $\sigma^2 = 1/(1-2\eta)$ and define
$\bar{Q}_{\mu,\sigma^2}$ as $N(-\mu/(2\sigma^2-1), \sigma^2/(2\sigma^2-1))$ with $\sigma^2 > 1/2$.
Applying Theorem~\ref{thm:theta*}, $(\mu^*,\sigma^*)$ is the root of 
\begin{eqnarray}
  \label{eq:tilting-param-normal}
\mu = E_{\bar{Q}_{\mu,\sigma^2}}[X|X>a]~~~{\rm and}~~~\sigma^2+\mu^2 = E_{\bar{Q}_{\mu,\sigma^2}}[X^2|X>a].
\end{eqnarray}


Take the one-parameter exponential embedding case, with $\sigma$ fixed. Standard calculation gives 
$\Psi(\theta) =\text{e}^{\theta^2/2}$,
$\psi(\theta) = \theta^2/2$ and $\psi'(\theta) = \theta$. 
Using the fact that $X|\{X>a\}$ is a truncated normal
distribution with minimum value $a$ under $\bar{Q}$, $\theta^\ast$
must satisfy $\theta =\frac{\phi(a+\theta)}{1-\Phi(a+\theta)}-\theta$, cf.~\cite{FuhHu2004}.

Table~\ref{tb:IS-normal} presents numerical results for the normal
distribution. As demonstrated in the table, using the sufficient exponential
tilting for the simple event,
$\mathbbm{1}_{\{X>a\}}$, yields performance 2 to 5 times better 
than the one-parameter exponential tilting in terms of
variance reduction factors.

\begin{table}[h]
  \scriptsize
\centering
\begin{tabular}{ccrrr}
\toprule[.04cm]
$[X\sim N(0,1)]$ & & \multicolumn{3}{c}{Variance reduction factors}\\
\cmidrule{3-5}
$a$&Crude for $P(X>a)$&$\mu^*$&$\sigma^*$&$(\mu^*,\sigma^*)$\\
\midrule
1&1.566$\times 10^{-1}$&5&2&12\\
2&2.300$\times 10^{-2}$ &19&4&60\\
3&1.370$\times 10^{-3}$&222&35&617\\
4&3.000$\times 10^{-5}$&7,094&860&32,552\\
\bottomrule[.04cm]
\end{tabular}
\caption{\textbf{Sufficient exponential importance sampling for normal distribution}
\label{tb:IS-normal} 
}
\end{table}


We now proceed to a $d$-dimensional multivariate normal distribution.
Let $X=(X_1,\ldots,X_d)^{\intercal}$ be a random vector with the standard multivariate normal distribution, denoted by $N(0,\mathbb{I})$, with pdf
$\text{det}(2\pi\mathbb{I})^{-1/2}\,e^{-(1/2)\,x^{\intercal}\mathbb{I}^{-1}x},$ where
$\mathbb{I}$ is the identity matrix.
By using the sufficient exponential embedding in (\ref{tilt}), we have 
\begin{eqnarray}\label{tilt-multvariate-inormal}
  \frac{dQ_{\theta,\eta}/{d\mathcal{L}}}{dP/{d\mathcal{L}}} &=& \frac{\exp\{\theta^{\intercal} x + x^{\intercal}Mx\}}{E[\exp\{\theta^{\intercal} X + X^{\intercal}MX\}]} \nonumber \\
  &=& \frac{e^{\theta^{\intercal}x+x^{\intercal}Mx-{\frac{1}{2}(\theta^{\intercal}(\mathbb{I}-2M)^{-1}\theta)}}}{\sqrt{|(\mathbb{I}-2M)^{-1}|}},
\end{eqnarray}
where $|\cdot|$ denotes the determinant of a matrix, and 
\begin{equation*}
  M=\left(a_{ij}\right)\in\mathbb{R}^{d\times d},
\end{equation*}
with $a_{ij}=\eta_i$ for $i=j$ and $a_{ij}=\eta_{d+1}$ for $i\neq j$.
In this case, the tilting probability measure $Q_{\theta,\eta}$ is
$N((\mathbb{I}-2M)^{-1}\theta,(\mathbb{I}-2M)^{-1})$.

For the event of $\wp(X)=\mathbbm{1}_{\{X \in A\}}$, define
$\bar{Q}_{\theta,\eta}$ as 
$N((\mathbb{I}+2M)^{-1}(-\theta),(\mathbb{I}+2M)^{-1})$.
Similar to the above one-dimensional normal distribution, we consider the
standard parameterization by letting $\mu:=(\mathbb{I}-2M)^{-1}\theta$, $\Sigma
:= (\mathbb{I}-2M)^{-1}$, and define $\bar{Q}_{\mu,\Sigma}$ as 
$N(-(2\mathbb{I}-\Sigma^{-1})^{-1}(\Sigma^{-1})\mu,(2\mathbb{I}-\Sigma^{-1})^{-1})$.
Applying Theorem~\ref{thm:theta*}, $(\mu^*,\Sigma^*)$ is the root of
\begin{eqnarray}
  \label{eq:multi_mu}
  \mu &=& {E_{\bar{Q}_{\mu,\Sigma}}\left[X | X\in A\right]},\\
  \label{eq:multi_sigma}
  \mathcal{K}(\mu,\Sigma)& =& {E_{\bar{Q}_{\mu,\Sigma}}\left[X^{\intercal}(\nabla_{\eta_i}M) X | X\in A\right]}\,\,\,\,\,\,\text{for }i=1,2,\cdots,d+1,
\end{eqnarray}
where
\begin{eqnarray}
  \label{eq:eta_b}
  \nabla_{\eta_i} M & = & (b_{jk})\in \mathbb{R}^{d\times d}\,\,\,\, \text{for }i=1,2,\cdots,d+1 ,\\
  \label{eq:eta_d}
  \mathcal{K}(\mu,\Sigma) & = & \frac{1}{2}\Tr\left(-(\nabla_{\eta_i}\Sigma^{-1})(\Sigma)\right)-\frac{1}{2}\mu^{\intercal}(\nabla_{\eta_i}\Sigma^{-1})\mu.
\end{eqnarray}

Here in Equation~(\ref{eq:eta_b}), $\Tr(A)$ is the trace of matrix $A$; the
value of $b_{jk}$ is defined as follows: for each $i=1,\cdots,d$,
\begin{equation*} 
  b_{jk} = 
  \begin{cases} 
    1, & \text{if } i=j=k,\\ 
    0, & \text{otherwise},
  \end{cases}
\end{equation*} 
and for $i=d+1$, 
\begin{equation*} 
  b_{jk} = 
  \begin{cases} 
    1, & \text{if } i\neq k,\\ 
    0, & \text{otherwise}.
  \end{cases}
\end{equation*} 

\begin{remark}
The left-hand sides (LHSs) of (\ref{eq:multi_mu}) and (\ref{eq:multi_sigma}) are
derivatives of the cumulant function $\psi_{MN}(\theta,\eta):= \log(\sqrt{|(\mathbb{I}-2M)^{-1}|})+\frac{1}{2}(\theta^{\intercal}(\mathbb{I}-2M)^{-1}\theta)$ for a multivariate
normal with respect to the parameters $\theta$ and $\eta$, respectively.
Note that in the RHS of (\ref{eq:eta_d}), Jacobi's formula is adopted for
the derivative of the determinant of the matrix $(\mathbb{I}-2M)^{-1}$ and
$\nabla_{\eta_i}\Sigma^{-1}$ is  
\begin{equation*}
  \nabla_{\eta_i}\Sigma^{-1}=\nabla_{\eta_i}\left(\mathbb{I}-2M\right)=(m_{jk})\in \mathbb{R}^{d\times d}\,\,\,\,\text{for }i=1,2,\cdots,d+1,
\end{equation*}
where for each $i=1,\cdots,d$, 
\begin{equation*} 
  m_{jk} = 
  \begin{cases} 
    -2, & \text{if } i=j=k,\\ 
    0, & \text{otherwise},
  \end{cases}
\end{equation*} 
and for $i=d+1$, 
\begin{equation*} 
  m_{jk} = 
  \begin{cases} 
    -2, & \text{if } i\neq k,\\ 
    0, & \text{otherwise}.
  \end{cases}
\end{equation*} 

\end{remark}

Table~\ref{tb:IS-multivariate-normal} presents the numerical results for the
standard bivariate normal distribution.
As shown in the table, for three types of events $\mathbbm{1}_{\{ X_1+X_2>a\}}, \mathbbm{1}_{\{X_1>a,X_2>a\}}$, and $\mathbbm{1}_{\{X_1X_2>a,X_1>0,X_2>0\}}$, tilting different parameters results in different performance in variance reduction.
Although sometimes tilting the variance parameter or the correlation parameter alone provides poor performance, combining them to the mean
parameter tilting can yield 2 to 3 times better performance than one-parameter exponential tilting.

\begin{table}[h]
  \scriptsize
\centering
\begin{tabular}{cccrrrr}
\toprule[.04cm]
$[X\sim N_2(0,\mathbb{I})]$ & & & \multicolumn{4}{c}{Variance reduction factors}\\
\cmidrule{4-7}
& $k$&Crude&$\mu^*$&$\sigma^*$&$\rho^*$&$(\mu^*, \sigma^*,\rho^*)$\\
\midrule
\multirow{3}{*}{$P(X_1+X_2>a)$}
&3&1.663$\times 10^{-2}$&24&2&2&43\\
&4&2.400$\times 10^{-3}$&138&4&2&354\\
&5&1.800$\times 10^{-4}$&1,064&8&4&4,036\\
\midrule
\multirow{3}{*}{$P(X_1>a,X_2>a)$}
&1&2.532$\times 10^{-2}$&9&1&2&16\\
&1.5&4.600$\times 10^{-3}$&34&2&7&68\\
&2&5.800$\times 10^{-4}$&227&5&18&504\\
\midrule
\multirow{3}{*}{$P(X_1X_2>a,X_1>0,X_2>0)$}
&2&1.538$\times 10^{-2}$&21&2&3&46\\
&3&4.800$\times 10^{-3}$&57&2&3&145\\
&5&5.600$\times 10^{-4}$&425&5&3&1,213\\
\bottomrule[.04cm]
\end{tabular}
\caption{\textbf{Sufficient exponential importance sampling for standard bivariate normal distribution}
\label{tb:IS-multivariate-normal} 
}
\end{table}
\end{example}



\begin{example}\label{exp:gamma}
Gamma distribution.
 
Let $X$ be a random variable with a Gamma distribution, denoted by
$\texttt{Gamma}(\alpha,\beta)$, with pdf
$\frac{dP}{d\mathcal{L}}=(\beta^{\alpha}/\Gamma(\alpha))\, x^{\alpha-1}e^{-\beta x}$.
By using the sufficient exponential embedding in (\ref{tilt}) with
$h_1(x):=\log (x)$ and $h_2(x):=x$, we have 
\begin{eqnarray}\label{tilt-gamma}
  \frac{dQ_{\theta,\eta}/{d\mathcal{L}}}{dP/{d\mathcal{L}}} &=& \frac{\exp\{\theta h_1(x) + \eta h_2(x)\}}{E[\exp\{\theta h_1(X) + \eta h_2(X)\}]} \nonumber \\
&=& e^{x\eta+\theta\log(x)}\frac{\Gamma(\alpha)}{(1/\beta)^{-\alpha}(\beta-\eta)^{-\alpha-\theta}\Gamma(\alpha+\theta)}.
\end{eqnarray}
In this case, the tilting probability measure $Q_{\theta,\eta}$ is
$\texttt{Gamma}(\alpha+\theta, \beta-\eta)$.
For the event of $\wp(X)=\mathbbm{1}_{\{X > a\}}$ for $a > 0$, define
$\bar{Q}_{\theta,\eta}$ as $\texttt{Gamma}(\alpha-\theta, \beta+\eta)$.
Applying Theorem~\ref{thm:theta*}, $(\theta^*,\eta^*)$ is the root of 
\begin{eqnarray}
  \label{eq:tilting-param-gamma-2}
  -\log (\beta-\eta) + \Upsilon(\alpha+\theta)&=& E_{\bar{Q}_{\theta,\eta}}\left[\,\log\left(X\right)|X>a\right],\\
  \label{eq:tilting-param-gamma-1}
  \frac{\alpha+\theta}{\beta-\eta} &=& E_{\bar{Q}_{\theta,\eta}}\left[X|X>a\right],
\end{eqnarray}
where $\Upsilon(\alpha+\theta)$ is a digamma function equal to $\Gamma'(\alpha+\theta)/\Gamma(\alpha+\theta)$.

Table~\ref{tb:IS-gamma} presents the numerical results for the Gamma
distribution.
Note that the commonly used one-parameter exponential tilting involves a change
for the parameter $\beta$ only (i.e., changing $\beta$ to $\beta-\eta^*$) in
the case of the Gamma distribution.
However, observe that tilting the other parameter $\alpha$ (i.e.,
$\alpha\rightarrow\alpha+\theta^*$) for some cases yields 2 to 3 times better
performance than the one-parameter exponential tilting in terms of variance
reduction factors.
This is due to the constraint $-\theta<\alpha$ and $\eta<\beta$.
For instance, consider the case in which we only tilt one parameter, either
$\theta$ or $\eta$ as follows.
For the simple event $\mathbbm{1}_{\{X>a\}}$, we can either choose a parameter
$\eta$ such that $0<\eta<\beta$ or a parameter $\theta$ such that $\theta>0$ to
obtain a larger mean for the tilted Gamma distribution. In this case, it is easy to see that
$\theta$-tilting yields a larger search space for parameters and thus achieves
better performance than $\eta$-tilting.
Note that event $\mathbbm{1}_{\{1/X>a\}}$ shows the opposite effect.

\begin{table}[h]
  \scriptsize
\centering
\begin{tabular}{ccrrrccrrr}
\toprule[.04cm]
\multicolumn{10}{c}{$[X\sim \texttt{Gamma}(\alpha,\beta)]$}\\
\cmidrule{1-10}
 & & \multicolumn{3}{c}{Variance reduction factors}
& & & \multicolumn{3}{c}{Variance reduction factors}\\
\cmidrule{3-5}\cmidrule{8-10}
$a$&Crude for $P(X>a)$&$\theta^*$&$\eta^*$&$(\theta^*,\eta^*)$ & $a$ &Crude for $P(1/X>a)$ & $\theta^*$ & $\eta^*$ & $(\theta^*,\eta^*)$\\
\midrule
10&2.613$\times 10^{-1}$ &3 & 2&3 & 0.2 & 2.438$\times10^{-1}$ & 2 & 4 & 6\\
20&1.050$\times 10^{-2}$ & 46& 24 & 47 & 0.5 & 1.864$\times10^{-2}$ & 15 & 41 & 45\\
30&1.800$\times 10^{-4}$ &1,288 & 567 & 1,307 & 1.5 & 3.100$\times10^{-4}$ & 294 & 1,321& 1,744\\
35&3.000$\times 10^{-5}$ &11,788& 4,788 & 12,226 & 2.5 & 6.000$\times10^{-5}$ & 2,156 & 11,904 &11,939\\
\bottomrule[.04cm]
\end{tabular}
\caption{\textbf{Sufficient exponential importance sampling for the Gamma distribution}
\label{tb:IS-gamma} 
}
\end{table}

\end{example}

\begin{remark}
  Observed from the above examples, the proposed sufficient exponential
  tilting gains the improvement in terms of variance reduction in two aspects.
  First, in some cases, titling multiple parameters simultaneously via our
  sufficient exponential tilting algorithm greatly improves the variance
  reduction performance; for example, in the case of simulating simple moderate deviation rare events,
  for the normal distribution, although
  changing solely $\sigma$ or $\rho$ may not provide great help in variance
  reduction, changing them together with $\mu$ can achieve 3 to 4 times better
  performance than the traditional mean-shift one-parameter tilting method.
  Second, in some cases, tilting some other parameters results in better
  performance; for instance, for the gamma distribution, changing the shape
  parameter $\alpha$ leads to better performance for some rare events, though the
  traditional one-parameter tilting always changes the rate parameter $\beta$.
  Moreover, for the simple cases, the computational time of such a
  two-parameter tilting method is almost the same as the one of one-parameter
  tilting since our algorithm always converges with 3 or 4 iterations when
  locating optimal tilting parameters.
  Analyses for more complex mixture distributions regarding the
  computational cost are provided in Section~\ref{sec:time}.

  On the other hand, for other events of interest, e.g., $E(X\textbf{1}_{\{0<X<a\}})$, other than
  the presented rare event cases, the proposed sufficient exponential tilting
  method always provides better performance than the one-parameter tilting
  method because the optimal two-parameter tilting includes the solution of
  one-parameter tilting.
  For example, in the case of simulating $E(X\textbf{1}_{\{0<X<1\}})$ for the
  standard noraml distribution, tilting the standard deviation together with
  the mean via our method gains 10 times better variance reduction performance
  than tilting the mean only.
\end{remark}

\section{Importance sampling for portfolio loss}\label{sec:sampling}

For easy presentation, this section is divided into three parts.
We first introduce the sufficient exponential tilting for normal mixture
distributions in Section~\ref{sec:normal-mixture}.
Section \ref{sec:tilted} provides importance sampling of $\varrho(Z,W)$ defined in (\ref{hzw}).
Section~\ref{sec:algo} summarizes the importance sampling algorithm for calculating the probability of the portfolio loss $P(L_n > \tau)$.

\subsection{Sufficient exponential tilting for normal mixture distributions}\label{sec:normal-mixture}

Recall that in Equation~(\ref{eq:Xk}), the latent random vector $X$ follows a
multivariate normal mixture distribution. 
In this section, for simplicity, we consider a one-dimensional normal mixture
distribution as an example to demonstrate the proposed sufficient 
exponential tilting.
Let $X$ be a one-dimensional normal mixture random variable with only one factor (i.e., $d=1$) such that
\begin{equation}
  X=\xi\sqrt{W}Z,
\end{equation}
where $\xi\in\mathbb{R}$, $Z\sim N(0,1)$, and $W$ is a non-negative and scalar-valued random 
variable which is independent of $Z$.
Since the random variable $W$ is independent of $Z$, by using the sufficient 
exponential embedding with $h_1(z)$ and $h_2(z)$ for $Z$ and $\tilde{h}_1(w)$ and $\tilde{h}_2(w)$ for $W$, we have
\begin{eqnarray}\label{eq:tilt-normal-mixture}
  \frac{dQ_{\theta_1,\eta_1,\theta_2,\eta_2}/{d\mathcal{L}}}{dP/{d\mathcal{L}}} &=& \frac{\exp\{\theta_1 h_1(z) + \eta_1 h_2(z)\}}{E[\exp\{\theta_1 h_1(Z) + \eta_1 h_2(Z)]\}} \frac{\exp\{\theta_2 \tilde{h}_1(w) + \eta_2 \tilde{h}_2(w)\}}{E[\exp\{\theta_2 \tilde{h}_1(W) + \eta_2 \tilde{h}_2(W)]\}},
\end{eqnarray}
where $\theta_1,\eta_1$ are the tilting parameters for $Z$ and
$\theta_2,\eta_2$ are the tilting parameters for $W$.

Now, we consider the standard parameterization by letting $\mu:=\theta_1/(1-2\eta_1), 
\sigma^2 := 1/(1-2\eta_1)$, $\theta:=\theta_2$, and $\eta:=\eta_2$ adopted in
the examples of Section~\ref{sec:is}.
If $W$ follows a Gamma distribution, $\texttt{Gamma}(\alpha,\beta)$, from
Equations~(\ref{tilt-normal}) and (\ref{tilt-gamma}),
Equation~(\ref{eq:tilt-normal-mixture}) becomes
\begin{eqnarray}\label{eq:tilt-normal-mixture1}
&~&  \frac{dQ_{\mu,\sigma,\theta,\eta}/{d\mathcal{L}}}{dP/{d\mathcal{L}}} \\ 
&= & \frac{1}{\sigma} \exp\{\mu z - \mu^2/2 + \frac{\sigma^2-1}{2\sigma^2} (z- \mu)^2\}\times
 e^{w\eta+\theta\log(w)}\frac{\Gamma(\alpha)}{(1/\beta)^{-\alpha}(\beta-\eta)^{-\alpha-\theta}\Gamma(\alpha+\theta)}.\nonumber
\end{eqnarray}
The optimal tilting parameters $\mu^*, \sigma^*$ for $Z$ can be obtained by
solving Equation~(\ref{eq:tilting-param-normal}), and $\theta^*$ and $\eta^*$
for $W$ are the solutions of~(\ref{eq:tilting-param-gamma-1})
and~(\ref{eq:tilting-param-gamma-2}).

Table~\ref{tb:IS-normal-mixture} tabulates the numerical results for this
one-dimensional, one-factor normal mixture distribution. Since 
for a normal mixture random variable, the variance is associated with the
random variable $W$, tilting the standard deviation $\sigma$ with
the mean $\mu$ of the normal random variable $Z$ is relatively insignificant in comparison to
tilting the parameters of $W$ (i.e., $\theta$ or $\eta$) with $\mu$. This is
shown in Table \ref{tb:IS-normal-mixture}. In addition, similar to the case demonstrated in Example~\ref{exp:gamma}, in this case, tilting $\theta$ with $\mu$ also yields better performance than
tilting $\eta$ with $\mu$ in the sense of variance reduction.

\begin{table}[h]
  \scriptsize
\centering
\begin{tabular}{cccrrrrrrr}
\toprule[.04cm]
$[Z\sim N(0,1), W\sim\texttt{Gamma}(\alpha,\beta)]$ & & & \multicolumn{7}{c}{Variance reduction factors}\\
\cmidrule{4-10}
& $a$&Crude&$\mu^*$&$\sigma^*$& $(\mu^*,\sigma^*)$&$\theta^*$&$\eta^*$& $(\mu^*,\theta^*)$&$(\mu^*,\eta^*)$\\
\midrule
\multirow{4}{*}{$P(\sqrt{W}Z>a)$}
&2&1.344$\times 10^{-1}$&3&1 & 6 &1&1&5&4\\
&4&2.808$\times 10^{-2}$&7&2& 10& 2&1&17&13\\
&8&9.500$\times 10^{-4}$&30&8& 48& 4&3&394&234\\
&12&2.000$\times 10^{-5}$&70&28 & 199 &11&4&9,619&5,054\\
\bottomrule[.04cm]
\end{tabular}
\caption{\textbf{Sufficient exponential importance sampling for normal mixture distribution}
\label{tb:IS-normal-mixture} 
}
\end{table}

Next, we summarize tilting for the event that the $k$-th obligor defaults if
$X_k$ exceeds a given threshold $\chi_k$ as ``\texttt{ABC}-event tilting'',
which actually involves the calculation of tail event
\begin{equation}
  \left\{(\texttt{A}+\texttt{B})\,{\texttt{C}}>\tau\right\},
\end{equation}
where \texttt{A} denotes the normal distributed part of the systematic risk
factors, \texttt{B} denotes the idiosyncratic risk associated with each
obligor, and \texttt{C} denotes the non-negative and scalar-valued random
variables which are independent of \texttt{A} and \texttt{B}.\footnote{Here we
  omit the coefficients before \texttt{A}, \texttt{B}, and \texttt{C} for
simplicity.}
For example, for the normal mixture copula model in (\ref{eq:Xk}), the
$d$-dimensional multivariate normal random vectors $Z=(Z_1,\cdots,Z_d)$ are
associated with \texttt{A}, the $\epsilon_k$ is associated with \texttt{B}, and
the non-negative and scalar-valued random variables $W$ are associated with
\texttt{C}.
Table~\ref{tb:XYZ} summarizes the exponential tilting used
in~\cite{glasserman2005importance,bassamboo2008portfolio,
chan2010efficient,scott2015general}, and our paper.  Note
that~\cite{glasserman2005importance} consider the normal copula model so that
there is no need for tilting~\texttt{C}.
It is worth mentioning that
\cite{bassamboo2008portfolio,chan2010efficient,scott2015general} only consider
the one-dimensional $t$-distribution, while we consider the multi-dimensional
normal mixture distribution.
Moreover, except for the proposed model, the other four methods adopt the so
called one-parameter tilting.
For example, even though~\cite{scott2015general} consider the tilting of
\texttt{A} and \texttt{C}, which is same as our setting, only one parameter is
tilted for each of the two distributions (i.e., mean for normal distribution
and shape for Gamma distribution); in our method, however, the tilting
parameter can be either the shape or the rate parameter for the underlying
Gamma distribution, which results in a more efficient simulation.

\begin{table}
  \setlength{\tabcolsep}{1pt}
  \scriptsize
\centering
\begin{tabular}{cccccc}
\toprule[.04cm]
& \multicolumn{4}{c}{One-parameter tilting (traditional)} & Sufficient exponential tilting (proposed) \\
\cmidrule(lr){2-5} \cmidrule(lr){6-6} 
& \cite{glasserman2005importance} & \cite{bassamboo2008portfolio} & \cite{chan2010efficient} & \cite{scott2015general} & Our paper\\
\cmidrule(lr){2-2} \cmidrule(lr){3-5} \cmidrule(lr){6-6} 
& Multivariate normal dist. & \multicolumn{3}{c}{$t$-dist.} & Normal mixture dist.\\
\midrule
\texttt{A} & \cmark & \xmark & \cmark & \cmark & \cmark \\
\texttt{B} & \cmark & \xmark & \cmark & \xmark & \xmark\\
\texttt{C} & NA &\cmark & \xmark & \cmark & \cmark \\ 
\bottomrule[.04cm]
\end{tabular}
\caption{\textbf{\texttt{ABC}-event tilting}}
\label{tb:XYZ} 
\end{table}

\subsection{Exponentially tilting for $\varrho({Z},{W})$}\label{sec:tilted}
In this subsection, we use the same notation as in
Section~\ref{sec:is}. Let ${Z} = (Z_1,\ldots,Z_d)^{\intercal}$ be a
$d$-dimensional multivariate normal random variable with zero mean and identity
covariance matrix $\mathbb{I}$,\footnote{Although here we consider the identity
covariance matrix for simplicity, it is straightforward to extend this to any valid
covariance matrix $\Sigma$.} and denote ${W}=(W_1,\ldots,W_{d+1})^{\intercal}$
as non-negative scalar-valued random variables, which are independent of $Z$. 
Under the probability measure $P$, let $f_1({ z})=f_1(z_1,\ldots,z_d)$ and
$f_2({w})=f_2(w_1,\ldots,w_{d+1})$ be the probability density functions of $Z$
and $W$, respectively, with respect to the Lebesgue measure $\mathcal{L}$.
As alluded to earlier, our aim is to calculate the expectation of $\varrho({Z},{W})$,
\begin{equation}
    \label{eq:m}
m=E_P\left[\varrho({Z},{W})\right]
\end{equation}
under the probability measure $P$.\footnote{Note that Theorem~\ref{thm:theta*} works for a random variable from an
exponential family. In this section, we simply apply the proposed importance sampling to simulate the
portfolio loss. In the case of simulating a rare event probability, to
which $X$ is in a non-convex set, further decomposition based on mixture
distribution tilting is required, cf. \cite{FuhHu2004,glasserman2008fast}.
Further study along this line will be studied in a separate paper.}

To evaluate (\ref{eq:m}) via importance sampling, we choose a sampling
probability measure $Q$, under which ${Z}$ and ${W}$ have  the corresponding
probability density functions $q_1({z})=q_1(z_1,\ldots,z_d)$ and
$q_2({w})=q_2(w_1,\ldots,w_{d+1})$.
Assume that $Q$ is absolutely continuous with respect to $P$;
Equation~(\ref{eq:m}) can then be written as
\begin{eqnarray}
    E_P\left[\varrho({Z},{W})\right]
    \label{eq:EQ}
    = E_Q\left[\varrho(Z,W)\frac{f_1(Z)}{q_1(Z)}\frac{f_2(W)}{q_2(W)}\right].
\end{eqnarray}

Let $Q_{\mu,\Sigma,\theta,\eta}$ be the sufficient exponentially tilted
probability measure of $P$. Here the subscripts
$\mu=(\mu_1,\ldots,\mu_d)^{\intercal}$, and $\Sigma$, constructed via $\rho$
and $\sigma=(\sigma_1,\ldots,\sigma_d)^{\intercal}$, are the tilting parameters
for random vector $Z$, and $\theta=(\theta_1,\ldots,\theta_{d+1})^{\intercal}$
and
$\eta=(\eta_1,\ldots,\eta_{d+1})^{\intercal}$ are the tilting parameters for $W$.
Define the likelihood ratios
\begin{eqnarray}
    \label{eq:expq1}
    r_{1,\mu,\Sigma}(z) = \frac{f_1(z)}{q_{1,\mu,\Sigma}(z)},~~~
   {\rm and}~~~
    r_{2,\theta,\eta}(w) =\frac{f_2(w)}{q_{2,\theta,\eta}(w)},
\end{eqnarray}
where $q_{1,\mu,\Sigma}(z)$ and $q_{2,\theta,\eta}(w)$ denote the probability
density functions corresponding to $q_1(z)$ with tilting parameters $\mu$ and $\Sigma$ and
$q_2(w)$ with tilting parameters $\theta$ and $\eta$, respectively.
Then, combined with (\ref{eq:expq1}), Equation~(\ref{eq:EQ}) becomes
\begin{equation}
  E_Q\left[\varrho(Z,W)\frac{f_1(Z)}{q_1(Z)}\frac{f_2(W)}{q_2(W)}\right]=E_{Q_{\mu,\Sigma,\theta,\eta}}\left[\varrho(Z,W){r_{1,\mu,\Sigma}(Z)}{r_{2,\theta,\eta}(W)}\right].
\end{equation}
Denote
\begin{eqnarray}
     \label{eq:opt}
G(\mu,\Sigma, \theta,\eta) = E_{P}\left[\varrho^2(Z,W){r_{1,\mu,\Sigma}(Z)}{r_{2,\theta,\eta}(W)}\right].
\end{eqnarray}
By using the same argument as that in Section \ref{sec:is}, we
minimize $G(\mu,\Sigma,\theta,\eta)$ to get the tilting formula. That is, 
tilting parameters $\mu^*$, $\Sigma^*$, $\theta^*$, and $\eta^*$ are chosen to 
satisfy
\begin{eqnarray}
    \label{eq:GZ}
    \frac{\partial G(\mu,\Sigma,\theta,\eta)}{\partial \mu}=0, &\,\,\,\,&\frac{\partial G(\mu,\Sigma,\theta,\eta)}{\partial \Sigma}=0,\\
    \label{eq:GW}
    \frac{\partial G(\mu,\Sigma,\theta,\eta)}{\partial \theta}=0, &\,\,\,\,&\frac{\partial G(\mu,\Sigma,\theta,\eta)}{\partial \eta}=0.
\end{eqnarray}

Note that to simulate the portfolio loss considered in (\ref{eq:loss}), 
$\varrho(z,w) = P(L_n>\tau| Z=z, W=w)$.
Therefore, to get the optimal tilting parameters in (\ref{eq:GZ}) and
(\ref{eq:GW}), we must calculate the conditional default probability $P(L_n>\tau| Z=z, W=w)$. 
We thus apply the fast inverse Fourier transform (FFT) method
described as follows.
First, though, we note that using the definition of the latent factor $X_k$
for the $k$-th obligor in Equation~(\ref{eq:Xk}), the conditional default
probability $P(X_k > \chi_k| Z=z, W=w)$ given $Z=(z_1,\ldots,z_d)^\intercal$
and $W=(w_1,\ldots,w_{d+1})^\intercal$ becomes
\begin{equation}
  \label{eq:marginalp}
  p_{z,w,k}=P\left(\left.\epsilon_k>\frac{\chi_k-\sum_{i=1}^d \rho_{ki}\sqrt{W_i}Z_i}{\rho_k\sqrt{W_{d+1}}}\,\,\right| Z=z, W=w\right).
\end{equation}

\subsubsection*{The (fast) inverse Fourier transform for non-identical $c_k$.}

With non-identical $c_k$, the distribution of the sum of $n$ independent but
non-identically distributed ``weighted'' Bernoulli random variables becomes
difficult to evaluate. Here we adopt the inverse Fourier transform to calculate
$\varrho(z,w)$~\citep{oberhettinger2014fourier}.
Recall that $L_n|(Z=z,W=w)$ equals
\begin{equation}
  L^{z,w}_n= \sum_{\ell=1}^n c_{\ell} H_{\ell}^{z,w},
\end{equation}
where $H_{\ell}^{z,w}\sim\texttt{Bernoulli}(p_{z,w,\ell})$, and the support of
$L^{z,w}_n$ is a discrete set with finite number of values.
Its Fourier transform is
\begin{equation}
  \phi_{L^{z,w}_n}(t)=E[e^{itL^{z,w}_n}]=E[e^{it\left(\sum_{i=1}^n c_{\ell} H^{z,w}_{\ell}\right)}]=\prod_{\ell=1}^n E[e^{it c_{\ell} H^{z,w}_{\ell}}]=\prod_{\ell=1}^n \phi_{H^{z,w}_{\ell}}(tc_{\ell}),
\end{equation}
where $\phi_{H^{z,w}_{\ell}}(s)=1-p_{z,w,\ell}+p_{z,w,\ell}e^{s}$.
For random variable $L^{z,w}_n$, we can recover $q^{z,w}_k$ = $P(L^{z,w}_n = k)$
by inverting the Fourier series:
\begin{equation}
  \label{eq:inversef}
  q^{z,w}_k=\frac{1}{2\pi}\int_{-\pi}^{\pi}e^{ikt}\prod_{\ell=1}^n\phi_{H^{z,w}_{\ell}}(tc_{\ell})dt,
\end{equation}
where $k=1,2,\cdots,\infty$.

A FFT algorithm computes the discrete Fourier transform
(DFT) of a sequence, or its inverse.
To reduce the computational time, this paper uses the FFT to approximate the
probability in Equation~(\ref{eq:inversef}).
With Euler's relation $e^{i\theta}=\cos\theta+i\sin\theta$, we can confirm 
that $\phi_{L^{z,w}_n}(t)$ has a period of $2\pi$; i.e.,
$\phi_{L^{z,w}_n}(t)=\phi_{L^{z,w}_n}(t+2\pi)$ for all $t$, which is due to the
fact that $e^{i(t+2\pi)k}=e^{itk}$.
With this periodic property, we now evaluate the characteristic function
$\phi_{L^{z,w}_n}$ at $N$ equally spaced values in the interval $[0,2\pi]$ as
\begin{equation*}
  b^{z,w}_m=\phi_{L^{z,w}_n}\left(\frac{2\pi m}{N}\right),\,\,\,m=0,1,\cdots,N-1,
\end{equation*}
which defines the DFT of the sequence of probabilities $q^{z,w}_k$.
By using the corresponding sequence of characteristic function values above, we
can recover the sequence of probabilities; that is, we aim for the sequence of
$q^{z,w}_k$'s from the sequence of $b^{z,w}_m$'s, which can be achieved by
employing the inverse DFT operation
\begin{equation}
  \tilde{q}^{z,w}_k=\frac{1}{N}\sum_{m=0}^{N-1}b^{z,w}_m e^{-i2\pi km/N},\,\,\,k=0,1,\cdots,N-1.
\end{equation}
Finally, the approximation of $\varrho(z,w)$ can be calculated as
\begin{equation}
  \label{eq:hzw2}
  \tilde{\varrho}(z,w)=1-P_{\text{FFT}}(L^{z,w}_n\leq\tau)=1-\sum_{\ell=0}^\tau \tilde{q}^{z,w}_{\ell},
\end{equation}
where $P_{\text{FFT}}(\cdot)$ denotes the probability approximated using a fast
inverse Fourier transform.

Note that even the ``exact'' FFT algorithms have errors when using
finite-precision floating-point arithmetic, but these errors are typically very
small.  Most FFT algorithms have outstanding numerical properties; for example,
the bound on the relative error for the Cooley-Tukey algorithm is $O(\epsilon
\log N)$. 
To attest the approximiation performance, Table~\ref{tb:fft} provides several
examples showing the approximation error and computational time of the inverse
Fourier transform.
In the table, we set the number of obligors $n=250$ and assume
$p_{z,w,\ell}=0.1$ (i.e., $H_{\ell}^{z,w}\sim\texttt{Bernoulli}(0.1)$)
for simplicity.
To check the approximation performance, we first consider the case with equal
$c_i=1$, where the probability (denoted as $P_{\text{Binomial}}(\cdot)$) is evaluated
analytically via the cumulative density function of the binomial distribution
with parameters $n=250$ and $p=0.1$.
Observe that the differences between the approximated probabilities
($P_{\text{FFT}}(\cdot)$) and the analytical ones
($P_{\text{Binomial}}(\cdot)$) are nanoscopic, i.e., the bias is extremely
small.
Moreover, we investigate the case with five different $c_i$, in which we
compare the approximated probabilities with the ones generated via simulation
with 500,000 samples; as shown in Table~\ref{tb:fft}, the approximated
probabilities all lie within the corresponding 95\% confidence intervals.
We note also that the computational time grows linearly with
the number of different $c_i$.\footnote{All of the experiments were obtained by
running programs via Mathematica 11 on a MacBook Pro with a 2.6 GHz Intel Core i7
CPU.}

\begin{table}[h]
  \scriptsize
\centering
\begin{tabular}{ccrcccccc}
\toprule[.04cm]
\multicolumn{4}{c}{$c_i=1$} & \multicolumn{4}{c}{$c_i =(\lceil 5i\rceil/n)^2$}\\
\cmidrule(lr){1-4} \cmidrule(lr){5-8}
$\tau$ & $P_{\text{FFT}}(L_n\leq\tau)$ & $P_{\text{FFT}}(L_n\leq\tau)-P_{\text{Binomial}}(L_n\leq\tau)$ & Time & $\tau$ & ${P}_{\text{FFT}}(L_n\leq \tau)$ &  $P_{\text{MC}}(L_n\leq\tau)$ (95\% CI) & Time\\
\midrule
20 & 1.72$\times 10^{-1}$ & $-$7.19$\times 10^{-15}$ & 0.03 & 200 & 1.29$\times 10^{-1}$ & 1.29$\times 10^{-1}$ (1.28$\times 10^{-1}$, 1.30$\times 10^{-1}$) & 0.13\\
10 & 3.53$\times 10^{-4}$ & $-$7.69$\times 10^{-17}$ & 0.04 & 100 & 1.32$\times 10^{-3}$ & 1.31$\times 10^{-3}$ (1.21$\times 10^{-3}$, 1.41$\times 10^{-3}$) & 0.14\\
5 & 5.84$\times 10^{-7}$  & 1.11$\times 10^{-16}$ & 0.04 & 50 & 1.20$\times 10^{-5}$ & 1.00$\times 10^{-6}$ (1.24$\times 10^{-6}$, 1.88$\times 10^{-5}$) & 0.14\\
\bottomrule[.04cm]
\end{tabular}
\caption{\textbf{Approximation performance and computational time (seconds) of the inverse Fourier transform}
\label{tb:fft} 
}
\end{table}

\subsection{Algorithms}\label{sec:algo}

This subsection summarizes the steps when we implement the proposed
sufficient exponential importance sampling algorithm, which consists of two components:
the tilting parameter search and the tail probability calculation.
The aim of the first component is to determine the optimal tilting parameters.
We implement the search phase using the automatic Newton's
method~\citep{teng2016automatic}.
We here define the conjugate measures
$\bar{Q}_{\mu,\Sigma}:=\bar{Q}^{\varrho(Z,W)}_{\mu,\Sigma}$,
$\bar{Q}_{\theta,\eta}:=\bar{Q}^{\varrho(Z,W)}_{\theta,\eta}$ of the measure Q with
respect to the payoff function $\varrho(Z,W)$.
With the conjugate measures and the results in (\ref{eq:multi_mu}),
(\ref{eq:multi_sigma}), (\ref{eq:tilting-param-gamma-1}), and
(\ref{eq:tilting-param-gamma-2}), we define functions $g_{\mu}(\mu)$,
$g_{\Sigma}(\Sigma)$, $g_{\theta}(\theta)$, $g_{\eta}(\eta)$ (see
Equations~(\ref{eq:GZ}) and (\ref{eq:GW})) as
\begin{eqnarray}
  \label{eq:g11}
  g_{\mu}(\mu) &=& \mu-{E_{\bar{Q}_{\mu,\Sigma}}\left[Z \,|\, L_n>\tau \right]},\\
  \label{eq:g12}
  g_{\Sigma}(\Sigma)&=&\mathcal{K}(\mu,\Sigma)-{E_{\bar{Q}_{\mu,\Sigma}}\left[Z^{\intercal}(\nabla_{\eta_i}M) Z \,|\, L_n>\tau\right]}\,\,\,\,\,\,\text{for }i=1,2,\cdots,d+1,\\
  \label{eq:g21}
  g_{\theta}(\theta)&=&
  \begin{bmatrix}
 -\log(\beta_1-\eta_1)+\Upsilon(\alpha_1+\theta_1),
    \cdots,
    -\log (\beta_{d+1}-\eta_{d+1})+\Upsilon(\alpha_{d+1}+\theta_{d+1})
\end{bmatrix}^{\intercal}\\
  &&-E_{\bar{Q}_{\theta,\eta}}\left[\,\ln\left(W\right) \,|\, L_n>\tau\right],\nonumber\\
  g_{\eta}(\eta)&=&
  \begin{bmatrix}
    \frac{\alpha_1+\theta_1}{\beta_1-\eta_1},
    \cdots,
    \frac{\alpha_{d+1}+\theta_{d+1}}{\beta_{d+1}-\eta_{d+1}}
\end{bmatrix}^{\intercal}
  -E_{\bar{Q}_{\theta,\eta}}\left[W \,|\, L_n>\tau\right],
  \label{eq:g22}
\end{eqnarray}
where $\nabla_{\eta_i}M$ in (\ref{eq:g12}) is defined in (\ref{eq:eta_b}).
To find the optimal tilting parameters, we must find the roots of the above
four equations.
With Newton's method, the roots of
(\ref{eq:g11}), (\ref{eq:g12}), (\ref{eq:g21}), and (\ref{eq:g22})
are found iteratively by
\begin{eqnarray}
  \label{eq:iterative}
  \delta^{(k)}&=&\delta^{(k-1)}-{J}^{-1}_{\delta^{(k-1)}}{g}_\delta(\delta^{(k-1)}),
\end{eqnarray}
where the Jacobian of $g_{\delta}(\delta)$ is defined as
\begin{equation}\label{jacobian}
  J_{\delta}[i,j]:= \frac{\partial}{\partial \delta_j} g_{\delta,i}(\delta).
\end{equation}
In (\ref{eq:iterative}) and (\ref{jacobian}) , $\delta$ can be replaced to $\mu$, $\Sigma$, $\theta$, and $\eta$, and
$J^{-1}_{\delta}$ is the inverse of the matrix $J_\delta$.

In order to measure the precision of roots to the solutions in
(\ref{eq:g11}), (\ref{eq:g12}), (\ref{eq:g21}), and
(\ref{eq:g22}), we define the sum of the square error of $g_\delta(\delta)$ as
\begin{equation}
\|g_\delta(\delta)\|=g'_\delta(\delta)g_\delta(\delta),
\end{equation}
and a $\delta^{(n)}$ is accepted when $\|g_\delta(\delta^{(n)})\|$ is less than
a predetermined precision level $\epsilon$.
The detailed procedures of the first component are described as follow:
\begin{itemize}
  \item
    Determine optimal tilting parameters:
    \begin{enumerate}[{(1)}]
      \item
        Generate independent samples $z^{(i)}$ from $N(0,\mathbb{I})$ and
        $q_j^{(i)}$ from $\texttt{Gamma}(\nu_j/2,1/2)$ for $i=1,\ldots \mathcal{B}_1$;
        calculate $w_j^{(i)}=\nu_j/q_j^{(i)}$ for $j=1,\dots,d+1$.
      \item
        Set $\mu^{(0)}$, $\Sigma^{(0)}$, $\theta^{(0)}$, and $\eta^{(0)}$
        properly and $k=1$.
      \item
        Calculate $g_{\delta^{(0)}}$ functions via
        (\ref{eq:g11}), (\ref{eq:g12}), 
 (\ref{eq:g21}), and
        (\ref{eq:g22}).
      \item
        Calculate the ${J}_\delta$ and their inverse matrices for the
        $g_\delta$ functions. 
      \item
        Calculate
        $\delta^{(k)}=\delta^{(k-1)}-{J}_{\delta^{(k-1)}}^{-1}{g}_\delta(\delta^{(k-1)})$
        in (\ref{eq:iterative}) for $\delta=\mu, \Sigma, \theta, \eta$. 
      \item
        Calculate $g_{\delta^{(k)}}$ functions via
        (\ref{eq:g11}), (\ref{eq:g12}),      
(\ref{eq:g21}), and
        (\ref{eq:g22}).  If $\forall\,\,\delta \in \{\mu,\Sigma,\theta,\eta\},\,\,
        \|{g}_\delta(\delta^{(k)})\|<\epsilon$, set $\delta^*=\delta^{(k)}$ and
        stop. Otherwise, return to step (4).
    \end{enumerate}
\end{itemize}

We proceed to describe the second component that calculates the probability of
losses, in which the optimal tilting parameters, $\mu^*$, $\Sigma^*$,
$\theta^*$ and $\eta^*$ are used (see the step (6) for the first component
above).
For demonstration purposes, we show the detailed procedure for
the $t$-copula model, in which the latent vector $X$ in Equation~(\ref{eq:Xk})
follows a multivariate $t$-distribution (recall that $W_j^{-1}=Q_j/\nu_j$ and
$Q_j\sim\texttt{Gamma}(\nu_j/2,1/2)$ for $j=1,\ldots,d+1$).
The detailed procedures of the second component are summarized as follows:
\begin{itemize}
  \item
    Calculate the probability of losses, $P(L_n>\tau)$:
    \begin{enumerate}[(1)]
      \item
        Generate independent samples $z^{(i)}$ from $N(\mu^*,\Sigma^*)$ and
        $q_j^{(i)}$ from $\texttt{Gamma}(\nu_j/2-\theta^*_j,1/2+\eta^*_j)$ for
        $i=1,\ldots \mathcal{B}_2$; calculate $w_j^{(i)}=\nu_j/q_j^{(i)}$ for
        $j=1,\dots,d+1$.
      \item
        Estimate $m$ by
        $\hat{m}=\frac{1}{\mathcal{B}_2}\sum_{i=1}^{\mathcal{B}_2}\tilde{\varrho}(z^{(i)},w^{(i)})\,r_{1,\mu^*,\Sigma^*}(z)\,r_{2,\theta^*,\eta^*}(w)$
        in (\ref{eq:expq1}), where $\tilde{\varrho}(z,w)$ is calculated by the
        analytical form from (\ref{eq:hzw2}) and
        $\mu^*,\Sigma^*,\theta^*,\eta^*$ is obtained from step (6) of the first
        component of the algorithm.
    \end{enumerate}
\end{itemize}

\begin{remark}
  As a side note, in the above sufficient exponential tilting algorithm, a
  componentwise Newton's method is adopted to determine the optimal tilting
  parameters $\mu,\Sigma,\theta,\eta$; this differs from Algorithm~2
  in~\cite{teng2016automatic}, which involves only one-parameter tilting for
  $\mu$.  
  It is worth mentioning that due the convex property of $G (\theta, \eta)$ and
  the uniqueness of the optimal tilting parameters, the optimal tilting formula is
  robust and not sensitive to the initial value. To illustrate this phenomenon,
  we consider the $G(\mu, \sigma, \rho)$ function for the simple event $X + Y> 3$
  under the standard bivariate normal distribution. As shown in
  Figure~\ref{fig:robust}, neither $\mu$ nor $\sigma$ are sensitive to the
  initial values; although the range of the initial values becomes vital for
  finding optimal $\rho$, $G(\rho)$ is flat for most $\rho$.
  Note that the variance reduction performance for this case is also shown in
  Table~\ref{tb:IS-multivariate-normal}.
  \begin{figure}
    \centering
    \begin{subfigure}[b]{0.3\textwidth}
      \includegraphics[width=\textwidth]{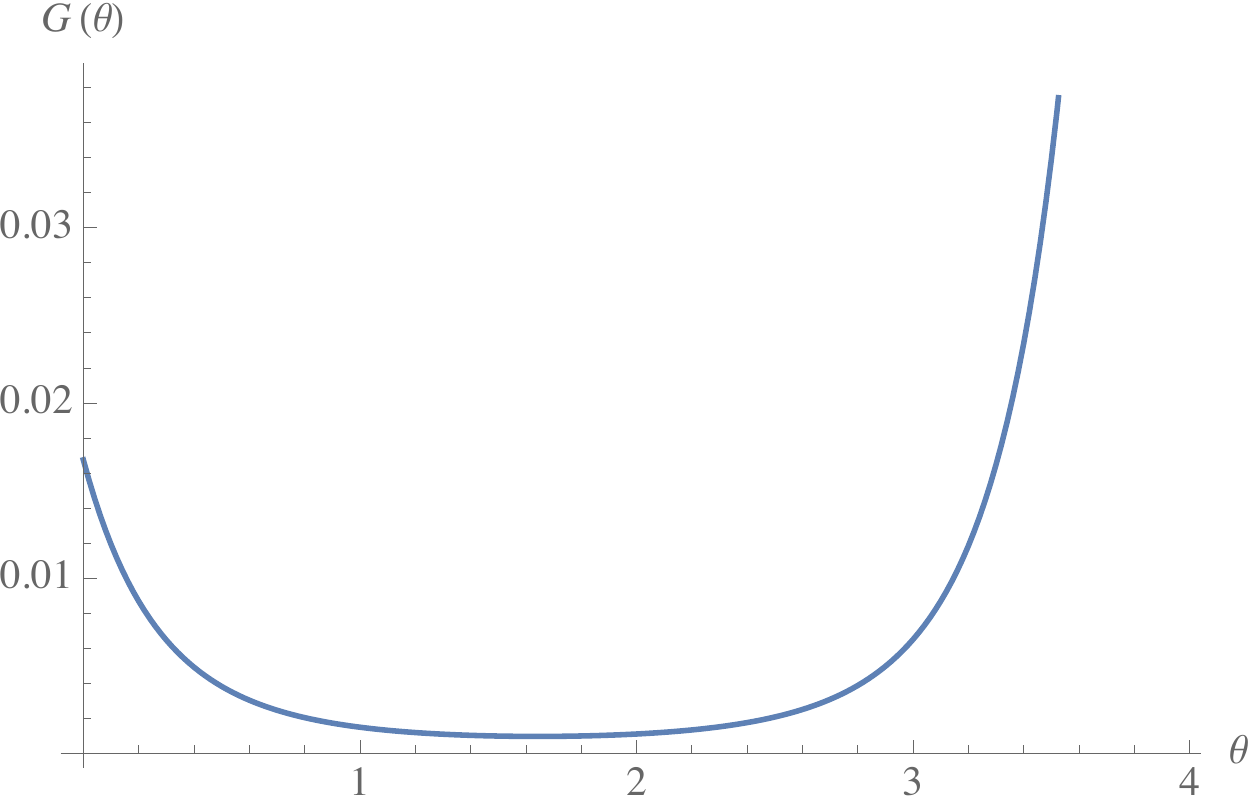}
      \caption{$G(\mu)$}
    \end{subfigure}
    ~ 
    \begin{subfigure}[b]{0.3\textwidth}
      \includegraphics[width=\textwidth]{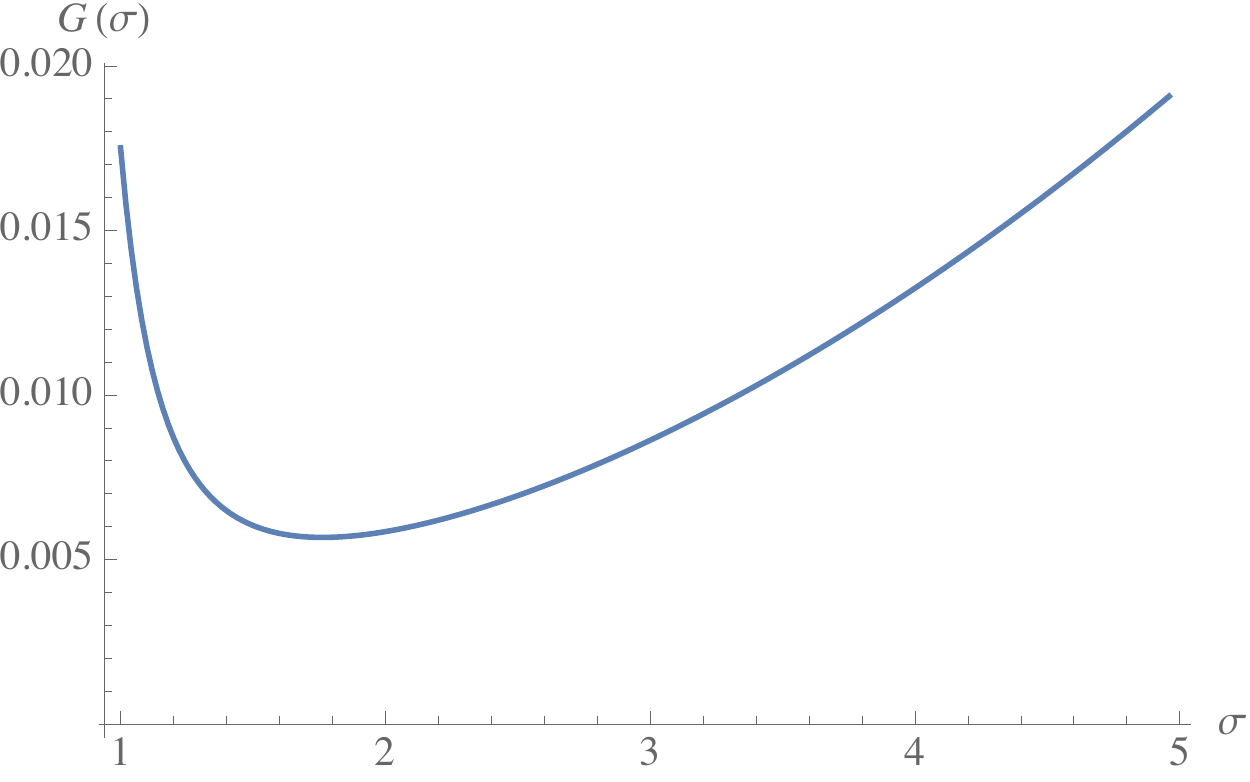}
      \caption{$G(\sigma)$}
    \end{subfigure}
    ~ 
    \begin{subfigure}[b]{0.3\textwidth}
      \includegraphics[width=\textwidth]{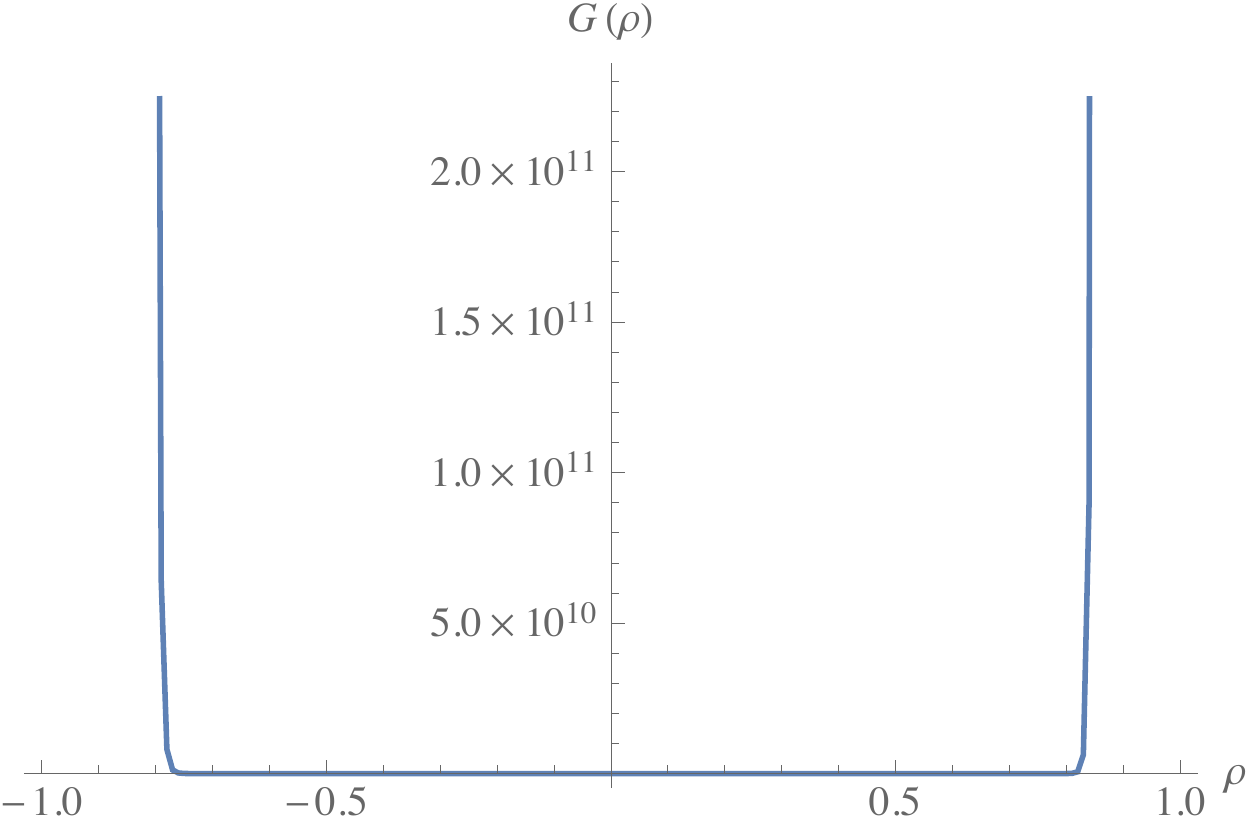}
      \caption{$G(\rho)$}
    \end{subfigure}
	 \caption{The $G(\mu,\sigma,\rho)$ function for the simple event $X+Y>3$
	 under the standard bivariate normal distribution}\label{fig:robust}
  \end{figure}
\end{remark}

\section{Numerical results}\label{sec:results}

This section demonstrates the capability and performance of the proposed method
via an extensive simulation study. 
We split the following discussion into four parts; the first three
are presented in Section~\ref{sec:numerical} and the last is presented in
Section~\ref{sec:time}.
First, for comparison purposes, we illustrate the results of the special case of
the normal mixture copula model, the $t$-copula model for single-factor
homogeneous portfolios, the settings of which are similar to those
in~\cite{bassamboo2008portfolio} and~\cite{chan2010efficient}.\footnote{Note
  that as stated in~\cite{scott2015general}, since their algorithm requires
  less computational time but is slightly less accurate
  than~\cite{chan2010efficient}, we here only compare the performance with
  ~\cite{bassamboo2008portfolio,chan2010efficient}.}
Second, we compare the performance of the proposed method with crude 
simulation, under 3-factor normal mixture models, in which the $t$-distribution
for $X_k$ is considered.
(Note that we only compare the performance of our method with crude Monte Carlo
simulation henceforth as most of the literature focuses on simulating
one-dimensional cases.)
Third, to evaluate the robustness of the proposed method, we also compare its performance
with that of crude simulation, under 3-factor normal mixture models, in which a
GIG distribution for $X_k$ is considered. The cases with different losses
resulting from default of the obligors are also investigated.
Finally, in Section~\ref{sec:time}, we compare the computational time of the
crude Monte Carlo simulation with the proposed importance sampling under several
scenarios, as well as provide insight into the trade-off between 
reduced variance and increased computational time.

Except for the experiments in Table~\ref{tb:time}, which compares the
computational time under different numbers of samples, we generate
$\mathcal{B}_1=5,000$ samples to locate the optimal tilting parameters and
$\mathcal{B}_2=10,000$ samples to calculate the probability of losses in all
of the rest experiments.
Following the settings in~\cite{bassamboo2008portfolio}, variances under crude
Monte Carlo simulation are estimated indirectly by exploiting the observation
that for a Bernoulli random variable with success probability $p$, the variance
equals $p(1-p)$.
Note that as mentioned in Section~\ref{sec:normal-mixture}, for
normal mixture random variables, tilting the variance of the normal random
variable $Z$ is relatively insignificant in comparison to tilting the parameters of $W$;
therefore, in the experiments, we only conduct mean tilting for $Z$ and
$\theta$- or $\eta$-tilting for $W$ for multi-factor normal mixture models.
It is worth mentioning that even in this case, our sufficient exponential importance
sampling method differs from previous studies, as only $\theta$ in the normal distribution and $\eta$ in the Gamma distribution are
considered as the tilting parameter in their methods. 

\subsection{Computation and numerical experiments on different model settings}\label{sec:numerical}
First, in Table~\ref{tb:onefactor} we compare the performance between our method
and the methods proposed in~\cite{bassamboo2008portfolio}
and~\cite{chan2010efficient}.
For comparison purposes, we adopt the same sets of parameter values as those in
Table 1 of~\cite{bassamboo2008portfolio}, where the latent variables $X_k$ in
Equation~(\ref{eq:Xk}) follow a $t$-distribution, i.e.,
$W_1^{-1}=W_2^{-1}=Q_1/\nu_1$ and $Q_1\sim\texttt{Gamma}(\nu_1/2,1/2)$.
The model parameters are chosen to be $n=250$, $\rho_{11}=0.25$, the default
thresholds for each individual obligors $\chi_i =0.5\times \sqrt{n}$, each
$c_i=1$, $\tau=250\times b$, $b=0.25$, and $\sigma_{\epsilon}=3$.
The table reports the results of the exponential change of measure (ECM) proposed
in~\cite{bassamboo2008portfolio} and conditional Monte Carlo simulation
without and with  cross-entropy (CondMC and CondMC-CE, respectively)
in~\cite{chan2010efficient}.
Observed from the table, the proposed algorithm (the last four columns) offers
substantial variance reduction compared with crude simulation, and in general,
it compares favorably to the ECM and CondMC estimators.
Moreover, in order to fairly compare with the results of CondMC-CE, we first
follow the CondMC method by integrating out the shock variable analytically;
then, instead of using the cross-entropy approach, we apply the proposed
importance sampling method for variance reduction.
Under this setting, the proposed method yields variance reductions comparable
to those of CondMC-CE.

\begin{table}
  \scriptsize
\centering
\begin{tabular}{ccrrrcrcr}
\toprule[.04cm]
&\multicolumn{2}{c}{\cite{bassamboo2008portfolio}} &\multicolumn{2}{c}{\cite{chan2010efficient}}&\multicolumn{4}{c}{Importance sampling (IS)}\\
\cmidrule(lr){2-3} \cmidrule(lr){4-5} \cmidrule(lr){6-9}
&&\multicolumn{1}{c}{ECM}&CondMC&CondMC-CE&\multicolumn{2}{c}{IS}&\multicolumn{2}{c}{CondMC-IS}\\
\cmidrule(lr){3-3} 
\cmidrule(lr){4-5} \cmidrule(lr){6-7} \cmidrule{8-9}
$\nu$ & $P(L_n>\tau)$ & V.R. factor & \multicolumn{2}{c}{V.R. factor} & $P(L_n>\tau)$  & V.R. factor & $P(L_n>\tau)$  & V.R. factor\\
\midrule
4&$8.13\times10^{-3}$&65&271&2,440&$8.11\times10^{-3}$&338&$8.09\times10^{-3}$&1,600\\
8&$2.42\times10^{-4}$&878&1,690&20,656&$2.36\times10^{-4}$&6,212&$2.47\times10^{-4}$&14,770\\
12&$1.07\times10^{-5}$&7,331&12,980&$2.08\times10^5$&$1.04\times10^{-5}$&16,100&$1.10\times10^{-5}$&$1.57\times10^5$\\
16&$6.16\times10^{-7}$&52,185&81,170&$1.30\times10^6$&$6.34\times10^{-7}$&$2.78\times10^5$&$6.20\times10^{-7}$&$1.89\times10^6$\\
20&$4.38\times10^{-8}$&301,000&$4.19\times10^5$&$1.27\times10^7$&$4.12\times10^{-8}$&$5.44\times10^6$&$4.14\times10^{-8}$&$1.61\times10^7$\\
\bottomrule[.04cm]
\end{tabular}
\caption{\textbf{Performance of proposed algorithm with equal loss resulting from
default of the obligors for a one-factor model ($t$-distribution)}
\label{tb:onefactor} 
}
\end{table}

Second, Tables~\ref{tb:equal-three-t}, \ref{tb:nonequal2-three-t},
and \ref{tb:nonequal5-three-t} compare the performance of the proposed importance
sampling method with crude simulation for a 3-factor $t$-copula model, a
special case of normal mixture copula models, where the latent variables $X_k$
follow a multivariate $t$-distribution, i.e., $W_j^{-1}=Q_j/\nu_j$ and
$Q_j\sim\texttt{Gamma}(\nu_j/2,1/2)$ for $j=1,\ldots,4$.
In the three tables, the results with equal losses
(Table~\ref{tb:equal-three-t}) and different losses
(Tables~\ref{tb:nonequal2-three-t} and~\ref{tb:nonequal5-three-t}) resulting
from the default of obligors with various parameter settings are listed.
Following the settings in~\cite{bassamboo2008portfolio}
and~\cite{chan2010efficient}, the threshold for the $i$-th obligor $\chi_i$ is
set to $0.5\times \sqrt{n}$, the idiosyncratic risk $\epsilon_k$ is set
to $N(0,9)$, and the total loss $\tau$ is set to $n\times b$.
The results with base-case model parameters are reported in the gray-background
cells in Tables~\ref{tb:equal-three-t}, \ref{tb:nonequal2-three-t}, and
\ref{tb:nonequal5-three-t}.
For the equal loss scenario ($c_i=1$), the model parameter $b$ for the base
case is set to 0.3, while for $c_i=(\lceil2i\rceil/n)^2$ and
$c_i=(\lceil5i\rceil/n)^2$, $b$ is set to 0.7 and 2, respectively.
In addition, the other parameters for the base case are listed as follows:
$\vec{\nu}=(8\,\,6\,\,4\,\,4)$, $n=250$, each $\rho_{ki}=0.1$ (for
$k=1,\ldots,n$ and $i=1,\ldots d$), and the covariance matrix
$\Sigma=(u_{ij})\in \mathbb{R}^{3\times 3}$ of the multivariate normal
distribution $Z$ is set to $u_{i,i} = \sigma_i^2,
u_{i,j}=u_{j,i}=\hat{\rho}\,\sigma_i\sigma_j$, where $\sigma_1=1, \sigma_2=0.8,
\sigma_3=0.5, \hat{\rho}=0.5$.

\begin{table}
  \scriptsize
\centering
\begin{tabular}{crrcrrcrr}
\toprule[.04cm]
\multicolumn{9}{c}{Equal loss ($c_i=1$)}\\
\midrule
$b$ & $P(L_n>\tau)$  & V.R. factor & $\vec{\nu}$ & $P(L_n>\tau)$  & V.R. factor & $n$ & $P(L_n>\tau)$ & V.R. factor\\
\cmidrule(lr){1-3}\cmidrule(lr){4-6}\cmidrule(lr){7-9}
\cellcolor{gray!25}{0.3}&	\cellcolor{gray!25}{3.08$\times10^{-3}$}&	\cellcolor{gray!25}{863}&	(4,4,4,4)&	3.09$\times10^{-3}$&	1,009&	100&	1.91$\times10^{-2}$&	416\\
0.4&	2.39$\times10^{-4}$&	5,931&	\cellcolor{gray!25}{(8,6,4,4)}&	\cellcolor{gray!25}{3.08$\times10^{-3}$}&\cellcolor{gray!25}{863}&\cellcolor{gray!25}{250}&	\cellcolor{gray!25}{3.08$\times10^{-3}$}&\cellcolor{gray!25}{863}\\
0.5&	2.13$\times10^{-6}$&	20,300&	(8,8,8,8)&	2.97$\times10^{-5}$&	1,667&	400&	1.17$\times10^{-3}$&	563\\
\midrule
$\rho_{ki}$ & $P(L_n>\tau)$  & V.R. factor & $\hat{\rho}$ & $P(L_n>\tau)$  & V.R. factor & $(\sigma_1,\sigma_2,\sigma_3)$ & $P(L_n>\tau)$ & V.R. factor\\
\cmidrule(lr){1-3}\cmidrule(lr){4-6}\cmidrule(lr){7-9}
\cellcolor{gray!25}{0.1}&	\cellcolor{gray!25}{3.08$\times10^{-3}$}&\cellcolor{gray!25}{863}&	-0.5&	3.06$\times10^{-3}$&	1,100&	(0.6,0.4,0.1)&	3.08$\times10^{-3}$&	991\\
0.3&	1.89$\times10^{-3}$&	945&	0&	3.05$\times10^{-3}$&	1,156&	(0.8,0.6,0.3) &	3.07$\times10^{-3}$&	1,087\\
0.5&	2.76$\times10^{-4}$&	1,174&	\cellcolor{gray!25}{0.5}&\cellcolor{gray!25}{3.08}$\times10^{-3}$&	\cellcolor{gray!25}{863}&\cellcolor{gray!25}{(1,0.8,0.5)} &\cellcolor{gray!25}{3.08$\times10^{-3}$}&\cellcolor{gray!25}{863}\\
\bottomrule[.04cm]
\end{tabular}
\caption{\textbf{Performance of proposed algorithm with equal loss resulting from
default of the obligors for a 3-factor model ($t$-distribution)}
\label{tb:equal-three-t} 
}
\end{table}

\begin{table}
  \scriptsize
\centering
\begin{tabular}{crrcrrcrr}
\toprule[.04cm]
\multicolumn{9}{c}{Two different losses ($c_i=(\lceil2i\rceil/n)^2)$}\\
\midrule
$b$ & $P(L_n>\tau)$  & V.R. factor & $\vec{\nu}$ & $P(L_n>\tau)$  & V.R. factor & $n$ & $P(L_n>\tau)$ & V.R. factor\\
\cmidrule(lr){1-3}\cmidrule(lr){4-6}\cmidrule(lr){7-9}
\cellcolor{gray!25}{0.7}&\cellcolor{gray!25}{4.79$\times10^{-3}$}&	\cellcolor{gray!25}{832}&	(4,4,4,4)&	4.78$\times10^{-3}$&	692&	100&	3.02$\times10^{-2}$&	325\\
1&	2.91$\times10^{-4}$&	3,078&	\cellcolor{gray!25}{(8,6,4,4)}&\cellcolor{gray!25}{4.79$\times10^{-3}$}&	\cellcolor{gray!25}{832}&	\cellcolor{gray!25}{250}&\cellcolor{gray!25}{4.79$\times10^{-3}$}&	\cellcolor{gray!25}{832}\\
1.2&	1.20$\times10^{-5}$&	14,471&	(8,8,8,8)&	8.64$\times10^{-5}$&	7,305&	400&	1.87$\times10^{-3}$&	739\\
\midrule
$\rho_{ki}$ & $P(L_n>\tau)$  & V.R. factor & $\hat{\rho}$ & $P(L_n>\tau)$  & V.R. factor & $\sigma_1,\sigma_2,\sigma_3$ & $P(L_n>\tau)$ & V.R. factor\\
\cmidrule(lr){1-3}\cmidrule(lr){4-6}\cmidrule(lr){7-9}
\cellcolor{gray!25}{0.1}&	\cellcolor{gray!25}{4.79$\times10^{-3}$}&	\cellcolor{gray!25}{832}&	-0.5&	4.81$\times10^{-3}$&	1,055&	(0.6,0.4,0.1)&	4.84$\times10^{-3}$&	700\\
0.3&	2.96$\times10^{-3}$&	876&	0&	4.80$\times10^{-3}$&	745&	(0.8,0.6,0.3) &	4.79$\times10^{-3}$&	582\\
0.5&	4.27$\times10^{-4}$&	739&	\cellcolor{gray!25}{0.5}&\cellcolor{gray!25}{4.79$\times10^{-3}$}&\cellcolor{gray!25}{832}&	\cellcolor{gray!25}{(1,0.8,0.5)} &\cellcolor{gray!25}{4.79$\times10^{-3}$}&\cellcolor{gray!25}{	832}\\
\bottomrule[.04cm]
\end{tabular}
\caption{\textbf{Performance of proposed algorithm with 2 different losses
resulting from default of the obligors for a 3-factor model with inverse FFT
($t$-distribution)}
\label{tb:nonequal2-three-t} 
}
\end{table}
As shown in Tables~\ref{tb:equal-three-t}, \ref{tb:nonequal2-three-t},
\ref{tb:nonequal5-three-t}, our IS approach performs significantly better than
crude simulation, especially when the loss threshold $\tau$ increases and the
probability becomes smaller.
Moreover, in contrast to the results listed in Table 2
of~\cite{chan2010efficient}, which show that when $\rho_{ki}$ increases, the
performance of CondMC deteriorates, the performance of our method is
demonstrated to be stable.
The reason for this phenomenon is due to the fact that as $\rho_{ki}$
increases, the factor $Z$ gains importance in determining the
occurrence of the rare event; while CondMC simply ignores the contribution of
$Z$, the proposed method twists the distributions of both $Z$ and $W$. 

Third, in addition to the $t$-copula model, Table~\ref{tb:equal-three-gig} shows the
results for another type of 3-factor normal mixture copula model, where $W_j$
follows a special case of generalized inverse Gaussian (GIG) distributions,
i.e., $W_j\sim\texttt{Gamma}(\nu_j/2,1/2)$ for $j=1,\ldots,4$ and
$\vec{\nu}=(8\,\,6\,\,4\,\,4)$; except for $b$ set to 0.28, 0.32, and 0.36, the
other model parameters are the same as those for the base case in
Table~\ref{tb:equal-three-t}.
From Table~\ref{tb:equal-three-gig} we observe that our approach outperforms
crude simulation, which attests the capability of the proposed
algorithm for normal mixture copula models.
Moreover, it is also worth noting that in this setting,
tilting the other parameter $\nu_j/2$ of the Gamma distribution (i.e.,
$\nu_j/2\rightarrow\nu_j/2-\theta^*_j$) yields 2 to 4 times better performance than
the traditional one-parameter exponential tilting (tilting $\eta$) in terms of the
variance reduction factors.\footnote{The phenomenon is consistent with the case
  demonstrated in Example~\ref{exp:gamma} and Section~\ref{sec:normal-mixture}.}


\begin{table}
  \scriptsize
\centering
\begin{tabular}{crrcrrcrr}
\toprule[.04cm]
\multicolumn{9}{c}{Five different losses ($c_i=(\lceil5i\rceil/n)^2)$}\\
\midrule
$b$ & $P(L_n>\tau)$  & V.R. factor & $\vec{\nu}$ & $P(L_n>\tau)$  & V.R. factor & $n$ & $P(L_n>\tau)$ & V.R. factor\\
\cmidrule(lr){1-3}\cmidrule(lr){4-6}\cmidrule(lr){7-9}
\cellcolor{gray!25}{2}&\cellcolor{gray!25}{2.38$\times10^{-2}$}&	\cellcolor{gray!25}{141}&	(4,4,4,4)&	2.39$\times10^{-2}$&	139&	100&	1.09$\times10^{-1}$&	59\\
4&	8.59$\times10^{-4}$&	3414&	\cellcolor{gray!25}{(8,6,4,4)}&\cellcolor{gray!25}{2.38$\times10^{-2}$}&\cellcolor{gray!25}{141}&	\cellcolor{gray!25}{250}&\cellcolor{gray!25}{2.38$\times10^{-2}$}&\cellcolor{gray!25}{141}\\
6&	4.15$\times10^{-7}$&	81,498&	(8,8,8,8)&	1.84$\times10^{-3}$&	1,131&	400&	9.77$\times10^{-3}$&	264\\
\midrule
$\rho_{ki}$ & $P(L_n>\tau)$  & V.R. factor & $\hat{\rho}$ & $P(L_n>\tau)$  & V.R. factor & $\sigma_1,\sigma_2,\sigma_3$ & $P(L_n>\tau)$ & V.R. factor\\
\cmidrule(lr){1-3}\cmidrule(lr){4-6}\cmidrule(lr){7-9}
\cellcolor{gray!25}{0.1}&\cellcolor{gray!25}{2.38$\times10^{-2}$}&\cellcolor{gray!25}{141}&	-0.5&	2.39$\times10^{-2}$&	152&	(0.6,0.4,0.1)&	2.37$\times10^{-2}$&	149\\
0.3&	1.51$\times10^{-2}$&	183&	0&	2.35$\times10^{-2}$&	125&	(0.8,0.6,0.3) &	2.40$\times10^{-2}$&	148\\
0.5&	2.34$\times10^{-3}$&	143&\cellcolor{gray!25}{0.5}&\cellcolor{gray!25}{2.38$\times10^{-2}$}&\cellcolor{gray!25}{141}&	\cellcolor{gray!25}{(1,0.8,0.5)} &\cellcolor{gray!25}{2.38$\times10^{-2}$}&\cellcolor{gray!25}{141}\\
\bottomrule[.04cm]
\end{tabular}
\caption{\textbf{Performance of our algorithm with 5 different losses resulting
from default of the obligors for a 3-factor model with inverse FFT ($t$-distribution)}
\label{tb:nonequal5-three-t} 
}
\end{table}
\begin{table}
  \scriptsize
\centering
\begin{tabular}{cccccrccr}
\toprule[.04cm]
&\multicolumn{2}{c}{Crude}&\multicolumn{3}{c}{IS (tilting $\eta$)}&\multicolumn{3}{c}{IS (tilting $\theta$)}\\
\cmidrule(lr){2-3} \cmidrule(lr){4-6} \cmidrule(lr){7-9} 
$b$& $P(L_n>\tau)$ & Variance & $P(L_n>\tau)$  & Variance & V.R. factor & $P(L_n>\tau)$  & Variance & V.R. factor\\
\midrule
0.28&	2.04$\times10^{-3}$&	2.04$\times10^{-3}$&		1.98$\times10^{-3}$&	6.99$\times10^{-6}$&	291&	1.98$\times10^{-3}$&	2.79$\times10^{-6}$&	731\\
0.32&	2.00$\times10^{-4}$&	2.00$\times10^{-4}$&		1.74$\times10^{-4}$&	8.09$\times10^{-8}$&	2,473&	1.75$\times10^{-4}$&	3.04$\times10^{-8}$&	6,575\\
0.36&	8.00$\times10^{-6}$&	8.00$\times10^{-6}$&		7.99$\times10^{-6}$&	3.77$\times10^{-10}$&	21,194&	7.55$\times10^{-6}$&	1.11$\times10^{-10}$&	72,352\\
\bottomrule[.04cm]
\end{tabular}
\caption{\textbf{Performance of our algorithm with equal loss resulting from
default of the obligors for a 3-factor model (symmetric generalized hyperbolic
distribution)}
\label{tb:equal-three-gig}
}
\end{table}

\subsection{Computational time of proposed importance sampling algorithm}\label{sec:time}
We now proceed to compare the computational time of crude Monte Carlo
simulation with the proposed importance sampling under several scenarios,
and provide insight into the trade-off between reduced variances
and increased computational time.
Table~\ref{tb:time} tabulates the computational time of crude simulation and
our method for the three cases listed in the top-left corner of
Table~\ref{tb:equal-three-t}.
From the table we observe that using more samples ($\mathcal{B}_1$) to determine
optimal tilting parameters greatly improves the variance reduction performance,
which however linearly increases the computational time to determine the
parameters; note that the variance reduction factor grows nonlinearly with the
number of samples $\mathcal{B}_1$, and our search algorithm generally
takes only 7 to 9 iterations to achieve convergence.\footnote{In all of the
  experiments reported here, the predetermined precision level
  $\epsilon$ is set to $10^{-4}$.}
Despite the need for additional computational time to find suitable
tilting parameters, we can use a mere $\mathcal{B}_2=1,000$ samples to obtain
rather good estimates with greatly reduced variances, especially when the tail
probability is small.
In the last column of Table~\ref{tb:time}, we also report the ratio of the
computational time consumed by the crude simulation generating a fair estimate
($\texttt{T}_C^*$, the value with a star symbol on the left-hand side) to that by our importance
sampling algorithm,
including the time for the parameter search ($\texttt{T}_I^1$) and probability calculation ($\texttt{T}_I^2$).
Observe from the table, for the case $b=0.5$, the crude simulation with
$10,000$ and even $100,000$ fails to generate the estimate, whereas our method
yields a good estimate with a 7,411 variance reduction ratio, while
the crude simulation requires 12.77 times more computational time than ours.
The relation between the variance reduction factor and the time consumption
ratio listed in the last two columns of Table~\ref{tb:time} suggests that the
proposed algorithm achieves good performance and thus makes a
practical contribution to measuring portfolio credit risk in normal mixture
copula models.

\begin{remark}
To find the optimal tilting parameters by solving the system of Eqs.
(\ref{eqn:foc1})-(\ref{eqn:foc2}), we have to evaluate the RHSs of Eqs.
(\ref{eqn:foc1})-(\ref{eqn:foc2}).  
Table~\ref{tb:time} shows that emprically the proposed recursive algorithm is
efficient in locating optimal tilting parameters as it always converges within
10 iterations with a rather small number of samples, e.g.,
$\mathcal{B}_1=1000$.
\end{remark}

\begin{table}[h]
  \setlength{\tabcolsep}{2.8pt}
  \scriptsize
\centering
\begin{tabular}{crrcrrccrlrr}
\toprule[.04cm]
& \multicolumn{3}{c}{Crude} & \multicolumn{8}{c}{IS }\\
\cmidrule(lr){2-4} \cmidrule(lr){5-12}
&&&& \multicolumn{3}{c}{Search parameters} & \multicolumn{5}{c}{Calculate probability ($\mathcal{B}_2=1,000$)}\\
\cmidrule(lr){5-7} \cmidrule(lr){8-12}
$b$ & $\mathcal{B}_2$ & Time ($\texttt{T}_C$) & $P(L_n>\tau)$ & $\mathcal{B}_1$ & Time ($\texttt{T}_I^1$) & \#iter &  Time ($\texttt{T}_I^2$)  & $P(L_n>\tau)$ & Variance & V.R. factor & $\texttt{T}_C^*/(\texttt{T}_I^1+\texttt{T}_I^2)$\\
\midrule
0.3 & 100 & 2 & -- & 100 & 191 & 9 & 200 & 3.20$\times 10^{-3}$ & 1.13$\times 10^{-3}$ & 3 & 0.46\\
 & 1,000 & 19 & 1.00$\times 10^{-3}$ & 500 & 647 & 8 & 191 & 3.38$\times 10^{-3}$ & 1.42$\times 10^{-4}$ & 22 & 0.22\\
 & 10,000 & {$^*$181} & 3.80$\times 10^{-3}$ & 1,000 & 2,049 & 8 & 201 & 3.01$\times 10^{-3}$ & 5.12$\times 10^{-6}$& 602 & 0.08\\
\midrule
0.4 & 1,000 & 17 & -- & 500 & 581 & 8 & 183 & 2.43$\times 10^{-4}$ & 1.37$\times 10^{-6}$ & 175 & 2.71\\
 & 10,000 & 184 & 1.00$\times 10^{-4}$ & 1,000 & 1,618 & 9 & 220 & 2.32$\times 10^{-4}$ & 8.08$\times 10^{-7}$ & 296 & 1.13\\
 & 100,000 & {$^*$2,070} & 2.00$\times 10^{-4}$ & 2,000 & 2,687 & 7 & 276 & 2.29$\times 10^{-4}$ & 2.83$\times 10^{-7}$ & 844 & 0.7\\
\midrule
0.5 & 10,000 & 204 & -- & 1,000 & 1,269 & 7 & 210 & 1.70$\times 10^{-6}$ & 2.88$\times 10^{-10}$ & 7,411 & 12.77\\
 & 100,000 & 1,812 & -- & 2,000 & 2,442 & 8 & 209 & 1.98$\times 10^{-6}$ & 2.59$\times 10^{-10}$ & 8,244 & 7.13\\
 & 1,000,000 & {$^*$18,896} & 2.00$\times 10^{-6}$ & 5,000 & 6,001 & 8 & 228 & 1.90$\times 10^{-6}$ & 1.07$\times 10^{-10}$ & 19,845 & 3.03\\
\bottomrule[.04cm]
\end{tabular}
\caption{\textbf{Computational times (seconds)}
\label{tb:time} 
}
\end{table}
\section{Efficient simulation for a multi-factor model on the CDXIG index}\label{sec:empirical}
To demonstrate the capability of the proposed method, in this section, we apply the 
importance sampling algorithm to a set of parameters of a multi-factor model
calibrated from data of the CDXIG index and consider the underlying
portfolio of the CDXIG index, using data from March~31,
2006~\citep{rosen2009valuing,rosen2010risk}. 
By using the same settings in \cite{rosen2010risk}, the multi-factor model is
assumed with a single global factor $Z_G$ and a set of sector factors $Z_{S_j}$
for $j=1,\cdots,7$; the details of the seven industrial sectors, merged from 25
Fitch sectors, can be found in~\cite{rosen2009valuing}.
For this eight-factor model,\footnote{Note that we treat this example as an
  eight-factor model; the method of combining our sufficient exponential tilting and the
  method proposed in~\cite{glasserman2008fast} will be a future work.} each
  obligor has two non-zero factor loadings, and the creditworthiness index of
  the $k$-th obligor in sector $S_j$ is thus given by
\begin{equation}\label{model_a}
  X_k=\sqrt{\rho_G}\cdot Z_G+ \sqrt{\rho_S-\rho_G} \cdot Z_{S_j} + \sqrt{1-\rho_S}\cdot\epsilon_k,\,\,\,\,\,\, \text{if } s(k)=S_j,
\end{equation}
where $s(k)$ denotes the sector for the $k$-th obligor, and $\rho_G=0.17$ and
$\rho_S=0.23$ are the same for all obligors.\footnote{These two values are
  chosen to match estimated correlations from~\cite{akhavein2005comparative}.}
As mentioned in~\citep{mcneil2015quantitative}, the latent variable $X_k$ can have
general interpretations, including asset value and creditworthiness.
For the creditworthiness in this example, the total portfolio loss is
thus defined as
\begin{equation}\label{eq:loss_a}
  L_n =c_1\mathbbm{1}_{\{X_1<\chi_1\}}+\cdots+c_n\mathbbm{1}_{\{X_n<\chi_n\}}.
\end{equation}
Since the CDXIG index has 125 equally-weighted obligors, we have $n=125$ and
set $c_1=c_2=\cdots=c_{125}$ equal to 1.\footnote{Note that here we follow the
  setting in~\cite{rosen2010risk} to define the the loss in (\ref{eq:loss_a}),
  which is different from that in (\ref{eq:loss}); the simulation scheme is thus
almost the same with a simple modification.}

Rather than using a conventional normal copula model in \cite{rosen2010risk}, 
here we consider this specification under a $t$-copula model.
The rational of using this model can be found in
\cite{frey2003dependent,bassamboo2008portfolio,mcneil2015quantitative}, in
which they claim that assuming a normal dependence structure may underestimate
the probability of joint large movements
of risk factors, while the $t$-copula model is better for modeling the effects
of extreme dependence.
For the $t$-copula model considered in (\ref{model_a}), 
the idiosyncratic risk $\epsilon_k$ is set to
$N(0,1)$, the threshold for the $i$-th obligor $\chi_i$ is set to $-0.55\times
\sqrt{n}$,\footnote{Since this multi-factor model is meant only to illustrate
  the methodology, for simplicity, the threshold for each obligor here is chosen to roughly
  match the average one-year default probability of the index, 0.19\%.}
the total loss $\tau$ is set to $n\times b$, the same setting as
in the previous subsection, and the latent variables $X_k$ follow a
multivariate $t$-distribution with a degree of freedom equal to 4, i.e.,
$W_j^{-1}=Q_j/4$ and $Q_j\sim\texttt{Gamma}(4/2,1/2)$ for $j=1,\ldots,4$.

\begin{table}
  \scriptsize
\centering
\begin{tabular}{ccrccr}
\toprule[.04cm]
&\multicolumn{2}{c}{Crude}&\multicolumn{3}{c}{IS}\\
\cmidrule(lr){2-3} \cmidrule(lr){4-6} 
$b$& $P(L_n>\tau)$  & Variance & $P(L_n>\tau)$  & Variance & V.R. factor\\
\midrule
0.01&		2.38$\times10^{-2}$&	2.32$\times10^{-2}$&		2.19$\times10^{-2}$&	2.61$\times10^{-4}$&		89\\
0.05&		6.60$\times10^{-3}$&	6.56$\times10^{-3}$&		6.43$\times10^{-3}$&	4.61$\times10^{-5}$&		142\\
0.2&		5.00$\times10^{-4}$&	5.00$\times10^{-4}$&		4.18$\times10^{-4}$&	1.01$\times10^{-6}$&		494\\
\bottomrule[.04cm]
\end{tabular}
\caption{\textbf{Performance of empirical example}
\label{tb:empirical}
}
\end{table}

Table~\ref{tb:empirical} tabulates the performance of this empirical example
in terms of variance reduction factors.
As shown in the table, the crude simulation suffers from the high variances of
the estimates, which results in extremely long simulation time to obtain a good
estimate, especially for very small probabilities.
For instance, for the case of $b=0.2$ in the table, we may need over 10,000
simulation paths to obtain a non-zero estimate for each simulation, resulting
in a large variance of $5.00\times10^{-4}$; hence to reduce variances and obtain a
good estimate, such simulations are necessarily lengthy.
For more efficient simulation schemes, although some previous studies
address this issue for the multi-factor model under normal copula
\citep{glasserman2008fast} or for the single-factor model under $t$-copula
\citep{bassamboo2008portfolio,chan2010efficient}, little work has been done 
on importance sampling methods for efficient simulation for multi-factor models
under the more general normal mixture copula model, which includes
the popular normal copula and $t$-copula models as special cases.
In particular, existing importance sampling algorithms cannot be used for
the portfolio loss in model (\ref{model_a}) and (\ref{eq:loss_a}).
To remedy this shortcoming, we propose an importance sampling algorithm to
estimate the probability that the portfolio incurs large losses under the
normal mixture copula.
It is worth mentioning again that the newly proposed sufficient exponential tilting
is more suitable for the normal mixture model simulations.
This is confirmed by the variance reduction performance of this method listed
in Table~\ref{tb:empirical} via the empirical example under an eight-factor
model.
Note that for this set of calibrated factor loadings, the idiosyncratic risk
with the rather high weight $\sqrt{1-\rho_S}\approx0.877$ dominates the value
of the latent variable $X_k$; in this case, tilting the last non-negative
scalar-valued random variable $W_{d+1}$ becomes critical for the variance
reduction performance.

For more applications, as future work we consider applying this method to estimate the
expected shortfall in more general copula models, such as Archimedean copula
models. Note that
the demand for simulation is even more serious in real applications, in
which the parameters in model (\ref{model_a}) are unknown, and must be
estimated from real data sets; see Chapter~11.5
of~\citep{mcneil2015quantitative}.
Moreover, when using the bootstrap method for accurate interval estimation for
the unknown parameters, a thousand replications of the bootstrap algorithm are
required.
We believe that the proposed importance sampling with modification, cf.
\cite{FuhHu2004}, would be useful in this setting.


\section{Conclusions}\label{sec:conclude}

This paper studies a multi-factor model with a normal mixture copula that allows
multivariate defaults to have an asymmetric distribution.
Due to the amount of the portfolio, the heterogeneous effect of obligors, and
the phenomena that default events are rare and mutually dependent, it  
is difficult to calculate portfolio credit risk either by means of
direct analysis or crude Monte Carlo simulation.
To address this problem, we first propose a general account of
a sufficient exponential tilting algorithm,
and then propose an efficient simulation algorithm to estimate the probability
that the portfolio incurs large losses under the normal mixture copula.
We also provide theoretical justifications of the proposed method and illustrate its
effectiveness through both numerical results and an empirical example.

There are several possible future directions based on this model. To name a
few, first, we will explore more properties of the proposed sufficient exponential tilting, and apply it to more
pratical cases, to see how far it can go. Second,
we may consider simulating portfolio loss under the meta-elliptical
copula and/or the Archimedean copula models. Third, although in this paper the
default time is fixed and the default boundary is exogenous, the default time
could be any time before a pre-fixed time $T$ and the default bounray could
depend on firm characteristic and may be state and time dependent.
To capture these phenomena, we will consider more complicated dynamic models, for which
the importance sampling should be more sophisticated.
Fourth, by developing the first-passage time model, it would be more practical
to consider the firm value process with credit rating. In such case, an
importance sampling for Markov chains is needed. \\

\noindent
{\bf Appendix: Proof of Theorem~\ref{thm:theta*}}

To prove Theorem~\ref{thm:theta*}, we require the following three propositions.
Proposition~\ref{pro2} is taken from Theorem~VI.3.4. of \cite{Ellis1985}, 
Proposition~\ref{pro1} is a 
standard result from convex analysis, and Proposition~\ref{pro3} is taken from
Theorem~1 of \cite{Soriano1994}. Note that although the function domain
is the whole space in
Propositions~\ref{pro1} and~\ref{pro3}, the results for a subspace still hold with similar proofs. 

\begin{proposition}\label{pro2}
 $f(\theta)$ is 
differentiable at $ \theta \in int(\Theta)$
if and only if the $d$ partial derivatives $\frac{\displaystyle \partial f(\theta)}{\displaystyle
\partial \theta_{i}}$  for $i =1,\cdots, d$ exist at $\theta \in int(\Theta)$ and are finite.
\end{proposition}

\begin{proposition}\label{pro1}
Let $f: \mathbbm{R}^d \to \mathbbm{R}$ be continuous on all of $x \in \mathbbm{R}^d$. If $f$ is coercive (in the 
sense that $f({\bf x}) \to \infty$ if $\|{\bf x}\| \to \infty$), then $f$ has at least one global 
minimizer.
\end{proposition}

\begin{proposition}\label{pro3}
Let $f: \mathbbm{R}^d \to \mathbbm{R}$, and $f$ is continuous differentiable function and convex, satisfies 
(\ref{minimizer}). Then a minimum point for $f$ exists.
\end{proposition}

\noindent
$<${\bf Proof of Theorem~\ref{thm:theta*}}$>$

In the following, we first show that $G(\theta,\eta)$ is a 
strictly convex function. For any given $\lambda \in(0,1)$, and $(\theta_1,\eta_1),
(\theta_2,\eta_2) \in  \Theta \times H$, by the convexity of $\psi(\cdot,\cdot)$, we have
\begin{eqnarray}\label{psi}
&~&\psi(\lambda \theta_1 + (1 - \lambda)\theta_2, \lambda \eta_1 + (1 - \lambda)\eta_2)
= \psi(\lambda (\theta_1, \eta_1) + (1 - \lambda) (\theta_2, \eta_2)) \\
&\leq& \lambda \psi(\theta_1, \eta_1) + (1 - \lambda) \psi(\theta_2, \eta_2). \nonumber
\end{eqnarray}
Then
\begin{eqnarray*}\label{convex}
&~& G(\lambda (\theta_1,\eta_1) + (1 - \lambda)(\theta_2,\eta_2))= G(\lambda \theta_1 + 
(1 - \lambda)\theta_2, \lambda \eta_1 + (1 - \lambda)\eta_2) \\ 
&=& E_p\bigg[\wp^2(X)\exp \big(-((\lambda \theta_1 + (1 - \lambda)\theta_2)^\intercal h_1(X) + (\lambda \eta_1 + (1 - \lambda)\eta_2)^\intercal h_2(X)) \\
&~&~~~~~~~ +\psi(\lambda \theta_1 + (1 - \lambda)\theta_2, \lambda \eta_1 + (1 - \lambda)\eta_2)\big) \bigg] \\
&\leq& E_p\bigg[\wp^2(X)\exp \big(-((\lambda \theta_1 + (1 - \lambda)\theta_2)^\intercal h_1(X) + (\lambda \eta_1 + (1 - \lambda)\eta_2)^\intercal h_2(X)) \\
&~& ~~~~ ~~~~ +\lambda \psi(\theta_1, \eta_1) + (1 - \lambda) \psi(\theta_2, \eta_2) \big) \bigg] ~~~{\rm by}~
(\ref{psi}) \\
&=& E_p\bigg[\wp^2(X)\exp \big(-\lambda (\theta_1^\intercal h_1(X) + \eta_1^\intercal h_2(X)) + \lambda \psi(\theta_1, \eta_1) - (1 - \lambda) (\theta_2^\intercal h_1(X) + \eta_2^\intercal h_2(X)) \\
&~&~~~~~~~~~ + (1 - \lambda) \psi(\theta_2, \eta_2) \big) \bigg] \\
&<& E_P\bigg[\lambda \wp^2(X)  e^{ - ( \theta_1^\intercal h_1(X) + \eta_1^\intercal h_2(X)) + \psi(\theta_1,\eta_1)}  +  (1 - \lambda) \wp^2(X)  e^{-( \theta_2^\intercal h_1(X) + \eta_2^\intercal h_2(X)) + \psi(\theta_2,\eta_2)}  \bigg]  \nonumber \\
&=& \lambda G(\theta_1,\eta_1) + (1 - \lambda )  G(\theta_2,\eta_2). \nonumber 
\end{eqnarray*}

Next, we prove the existence of $(\theta,\eta)$ in the optimization problem (\ref{eq:G}).
To get the global minimum of $G(\theta,\eta)$, we note that $G(\theta,\eta)$ is
strictly convex from the above argument, and $\frac{\displaystyle \partial
G(\theta,\eta)}{\displaystyle \partial \theta_{i}}$ and $\frac{\displaystyle
\partial G(\theta,\eta)}{\displaystyle \partial \eta_{i}}$ exists for 
$i =1,\cdots, p$, $j =1,\cdots, q$. Proposition \ref{pro2} establishes that
$G(\theta,\eta)$ is continuously differentiable for $(\theta,\eta) \in \Theta
\times H$. By the definition of $G(\theta,\eta)$ in (\ref{eq:G}), it is easy to
see that condition i) implies that $G(\theta)$ is coercive. Then by Proposition
\ref{pro1}, $G(\theta,\eta)$ has a unique minimizer. It is easy to see that ii)
implies conditions in Proposition \ref{pro3} hold. 

To prove (\ref{eq:theta_ast1}) and (\ref{eq:theta_ast2}), we simplify
the right-hand side of (\ref{eqn:foc1}) and (\ref{eqn:foc2}) under
$\bar{Q}_{\theta,\eta}$.
Standard algebra gives
\begin{eqnarray*}
\label{eq:ab_temp}
\frac{E_{P}\left[\wp^2(X)h_1(X) \mbox{e}^{-(\theta^\intercal h_1(X) + \eta^\intercal h_2 (X))}\right]}{E_{P}\left[\wp^2(X)\mbox{e}^{-(\theta^\intercal h_1(X) + \eta^\intercal h_2(X))}\right]}&=&E_{\bar{Q}_{\theta,\eta}}[h_1(X)], \\
 \frac{E_{P}\left[\wp^2(X) h_2(X)\mbox{e}^{-(\theta^\intercal h_1(X) + \eta^\intercal h_2(X))}\right]}{E_{P}\left[\wp^2(X)\mbox{e}^{-(\theta^\intercal h_1(X) + \eta^\intercal h_2(X))}\right]} 
 &=& E_{\bar{Q}_{\theta,\eta}}[ h_2(X)]
\end{eqnarray*}
for $i=1,\cdots,p,~j=1,\cdots,q$.
This implies the desired result. 
$\hfill \Box$~\\

\theendnotes

\ACKNOWLEDGMENT{
This research was partially supported by the Ministry of Science and Technology
in Taiwan under the grants MOST 105-2410-H-008-025-MY2, MOST
106-2118-M-008-002-MY2, MOST 102-2221-E-845-002-MY3, and MOST
105-2221-E-001-035.
}

\bibliographystyle{informs2014}
\bibliography{paper}
\end{document}